\documentclass[11pt,a4paper]{article}

\usepackage{amssymb}
\usepackage{graphicx}
\usepackage{bm}

\usepackage{mathrsfs}
\usepackage{cite}

\usepackage{amsmath}
\usepackage{microtype}

\allowdisplaybreaks

\newcommand*\xbar[1]{%
  \hbox{%
    \vbox{%
      \hrule height 0.5pt % The actual bar
      \kern0.5ex%         % Distance between bar and symbol
      \hbox{%
        \kern-0.1em%      % Shortening on the left side
        \ensuremath{#1}%
        \kern-0.1em%      % Shortening on the right side
      }%
    }%
  }%
}

\begin{document}

\title{Leading Yukawa terms in the three-loop anomalous dimension and four-loop $\beta$-function of ${\cal N}=1$ supersymmetric theories for various renormalization prescriptions}

\author{
R.R.Jumanov and K.V.Stepanyantz\\
\\
{\small{\em Moscow State University}}, {\small{\em  Faculty of Physics, Department  of Theoretical Physics}}\\
{\small{\em 119991, Moscow, Russia}}\\
}

\maketitle

\begin{abstract}
For an arbitrary renormalizable ${\cal N}=1$ supersymmetric theory with a single gauge coupling regularized by higher covariant derivatives we calculate the three-loop contribution to the anomalous dimension of the matter superfields proportional to the sixth powers of Yukawa couplings for a wide class of renormalization prescriptions. Next, taking into account that in the case of using the higher covariant derivatives supplemented by minimal subtractions of logarithms the NSVZ relation is valid in all orders, we calculate the corresponding four-loop contribution to the gauge $\beta$-function and investigate its scheme dependence. In particular, we construct the so-called ``minimal'' renormalization prescription for which the renormalization group functions have the simplest possible form provided that the exact equations for the gauge and Yukawa $\beta$-functions following from supersymmetry remain valid.
\end{abstract}

\section{Introduction}
\hspace*{\parindent}

Ultraviolet divergences in supersymmetric theories have a lot of interesting properties. For instance, even for ${\cal N}=1$ supersymmetric theories the gauge and Yukawa $\beta$-functions are related to the anomalous dimension of the matter superfields in all orders. For the Yukawa couplings, this statement follows from the nonrenormalization of the superpotential \cite{Grisaru:1979wc} and is rather evident in the superfield formalism. The equation relating the gauge $\beta$-function to the anomalous dimension is obtained in a more complicated way and is called ``the exact NSVZ $\beta$-function'' \cite{Novikov:1983uc,Jones:1983ip,Novikov:1985rd,Shifman:1986zi}. Some interesting features also exist in supersymmetric theories finite in the one-loop approximation \cite{Parkes:1984dh,Kazakov:1986bs,Ermushev:1986cu,Lucchesi:1987he,Lucchesi:1987ef}. Namely, in this case it is possible to tune a renormalization prescription in such a way as to obtain the all-loop finiteness \cite{Ermushev:1986cu}. In particular, if in a certain subtraction scheme the theory is finite in a given order, then its $\beta$-function vanishes in the next order \cite{Parkes:1985hj,Grisaru:1985tc}, see also \cite{Stepanyantz:2021dus}. Another interesting particular case is the so-called $P=\frac{1}{3} Q$ theories \cite{Jack:1995gm,Jack:1996qq}, for which the ratio of the Yukawa couplings to the gauge coupling is a renormalization group invariant in the first two orders \cite{Jack:1996qq}. This property hints at the possibility of reducing the number of independent couplings, see, e.g., \cite{Mondragon:2013aea,Heinemeyer:2014vxa,Heinemeyer:2019vbc}. However, in the next (three-loop) order the renormalization group invariance of the ratio $\lambda^{ijk}/e$ is broken by certain terms \cite{Jack:1996qq} proportional to $\zeta(3)$. Possibly, they could be eliminated by imposing more complicated constraints on the parameters of the theory, but the form of these constraints and even the very existence of them are not established at present. One can also suggest that the terms proportional to $\zeta(3)$ for a certain renormalization prescription come only from nonplanar superdiagrams \cite{Jack:2014pua,Kuzmichev:2023zxy}.

However, various exact equations relating the renormalization group functions (RGFs) of supersymmetric theories are scheme dependent. In particular, the scheme dependence of the NSVZ equation was investigated in \cite{Kutasov:2004xu,Kataev:2013csa,Kataev:2014gxa}. However, it was unclear for a long time for which renormalization prescription this equation is satisfied. In particular, it was demonstrated \cite{Jack:1996vg,Jack:1996cn,Jack:1998uj} that in the $\overline{\mbox{DR}}$ scheme (when a theory is regularized by dimensional reduction \cite{Siegel:1979wq} and divergences are removed by modified minimal subtraction \cite{Bardeen:1978yd}) the NSVZ relation is not valid starting from the three-loop approximation for the $\beta$-function and the two-loop approximation for the anomalous dimension. It should be noted that most calculations in the $\overline{\mbox{DR}}$ scheme were made using the manifestly supersymmetric supergraph technique. However, it is noticed \cite{Siegel:1980qs} that the regularization by dimensional reduction can produce some ambiguous expressions, which could break supersymmetry in higher orders as asserted in Refs. \cite{Avdeev:1982np,Avdeev:1982xy}. In particular, based on the calculation performed in components in Ref. \cite{Avdeev:1982np}, it was claimed   that supersymmetry becomes broken already in the three-loop approximation. In the subsequent paper \cite{Velizhanin:2008rw} a mistake in \cite{Avdeev:1982np} has been corrected, but for ${\cal N}=2$ supersymmetric Yang--Mills theory the problem still remained. However, the recent analysis made in \cite{Chakraborty:2026hdv} led to its solution and demonstrated that the dimensional reduction is supersymmetric, at least, in the three-loop approximation. Therefore, the fact that the NSVZ equation does not hold in the $\overline{\mbox{DR}}$ scheme is not related to the supposed supersymmetry breaking by dimensional reduction in higher orders. Certainly, this result exactly agrees with numerous investigations of its scheme dependence, e.g., \cite{Jack:1996vg,Jack:1996cn,Jack:1998uj,Kutasov:2004xu,Kataev:2013csa,Kazantsev:2020kfl} etc. Therefore, for the NSVZ equation to hold, not only supersymmetry but also a special renormalization prescription is required.

The problem of constructing the NSVZ scheme has been solved with the help of the Slavnov's higher covariant derivative regularization \cite{Slavnov:1971aw,Slavnov:1972sq,Slavnov:1977zf} (in the ${\cal N}=1$ superspace formulation \cite{Krivoshchekov:1978xg,West:1985jx}) in Refs. \cite{Stepanyantz:2016gtk,Stepanyantz:2019ihw,Stepanyantz:2020uke} (see \cite{Stepanyantz:2019lyo,Stepanyantz:2023jot} for review). Namely, the NSVZ equation is satisfied in all orders if a theory is regularized by higher covariant derivatives and divergences are removed by minimal subtractions of logarithms \cite{Kataev:2013eta,Stepanyantz:2020uke}.\footnote{In the Abelian case one more all-loop NSVZ renormalization prescription is the on-shell scheme \cite{Kataev:2019olb}.} The resulting renormalization prescription was called ``the HD+MSL scheme'' \cite{Stepanyantz:2017sqg,Shakhmanov:2017wji}. This statement has been confirmed by numerous calculations made in, e.g., Ref. \cite{Stepanyantz:2011jy,Kataev:2017qvk,Shakhmanov:2017soc,Kazantsev:2018nbl,Kuzmichev:2019ywn,Aleshin:2020gec,Shirokov:2023jya}. Moreover, if we know for what renormalization prescription the NSVZ equation is valid, then it is possible to considerably simplify the calculations of higher-order contributions to the $\beta$-function because the NSVZ equation relates the $\beta$-function in a certain order to the anomalous dimension of the matter superfields in all previous orders. This method was also frequently used for making multiloop calculations, see, e.g., \cite{Jack:1996cn,Jack:1998uj,Kazantsev:2020kfl,Haneychuk:2022qvu,Shirokov:2022jyd,Haneychuk:2025ejd}. Taking into account that the HD+MSL scheme is an all-loop renormalization prescription, we see that the higher covariant derivative regularization has certain significant advantages in comparison with other regularizations \cite{Jack:1997sr,Gnendiger:2017pys} in the supersymmetric case (besides the fact that it is formulated in the integer dimension).

Note that, using the higher covariant derivative regularization, it is also possible to construct the holomorphic coupling which is renormalized only in the one-loop approximation \cite{Shifman:1986zi}. For this purpose, according to \cite{Arkani-Hamed:1997qui}, it is necessary to make certain rescaling transformations. For theories regularized by higher covariant derivatives these rescaling transformations are equivalent to changing the masses of the Pauli--Villars superfields, see \cite{Stepanyantz:2004sg} for detail. In particular, the usual renormalization of the gauge coupling is obtained if their masses are proportional to the dimensionful cutoff parameter $\Lambda$ in the higher derivative term, for instance, $M=c\Lambda$ for ${\cal N}=1$ SQED, where $c$ is a constant independent of couplings. The one-loop renormalization of the holomorphic coupling (again, to be specific, for ${\cal N}=1$ SQED) appears for the mass $M=c\Lambda/Z$, where $Z$ is the renormalization constant for the matter superfields in any NSVZ scheme. Then divergences in the gauge sector will be present only in the one-loop approximation, because all higher order divergent contributions are in fact absorbed by the redefinition of the Pauli--Villars mass. (The importance of having two different regularization scales ($M$ and $\Lambda$) for investigating similar issues was discussed in \cite{Novikov:1985mf}.) Therefore, the higher covariant derivative regularization allows for explaining how to construct the holomorphic coupling and for which renormalization prescriptions it is not renormalized beyond the one-loop approximation.

However, the higher covariant derivative regularization is not uniquely defined. Really, the form of the higher derivative term may be different. It is also possible to choose different masses of the Pauli--Villars superfields which are used for regularizing the residual one-loop divergences \cite{Slavnov:1977zf}. Starting from a certain approximation (three loops for the gauge $\beta$-functions and two loops for the anomalous dimensions and the Yukawa $\beta$-functions) RGFs become dependent on a particular choice of the regularization parameters. Moreover, the renormalization procedure is not uniquely defined and leads to the appearance of some finite constants fixing a subtraction scheme in each order of the perturbation theory. Therefore, for an arbitrary renormalization prescription, RGFs start to depend on a large number of regularization and renormalization parameters in higher orders of the perturbation theory. Even the NSVZ schemes form the continuous set described by a certain number of free parameters \cite{Goriachuk:2018cac,Goriachuk_Conference}. That is why it is desirable to formulate a prescription how to fix all these arbitrary parameters. Certainly, in the supersymmetric case $\overline{\mbox{DR}}$ scheme allows solving this problem. However, the corresponding renormalization prescription cannot be considered the best because such a feature of supersymmetric theories as the NSVZ relation is not valid in this case. On the other side, the higher covariant derivative regularization produces such loop integrals that are very difficult to calculate. These integrals are especially complicated for the scheme dependent contributions to RGFs (certainly, in comparison with the integrals which determine scheme independent terms), see, e.g., \cite{Shirokov:2022jyd}. Therefore, we face the problem of constructing a special renormalization prescription for which RGFs have the simplest possible form, while all relations produced by supersymmetry remain valid. In Ref. \cite{Shirokov:2022jyd} such a ``minimal'' scheme has been constructed for ${\cal N}=1$ supersymmetric electrodynamics (SQED) with $N_f$ flavors in all orders. It appeared that in this scheme (at least, in the four-loop approximation for the $\beta$-function and three-loop anomalous dimension of the matter superfields) integrals obtained with the higher covariant derivative regularization do not depend on regularization parameters and can be calculated for an arbitrary choice of the higher derivative regulators. Taking into account that in this case all arbitrary renormalization parameters are fixed, this prescription seems very attractive. Certainly, it is desirable to generalize it to the (much more complicated) case of non-Abelian supersymmetric theories. For instance, in \cite{Haneychuk:2025ehb} the minimal scheme has been constructed for ${\cal N}=1$ SQCD+SQED in the three-loop approximation.

In this paper we consider an arbitrary ${\cal N}=1$ supersymmetric gauge theory with a single gauge coupling and calculate the three-loop contribution to the anomalous dimension of the matter superfields proportional to the sixth power of the Yukawa couplings in the case of using the higher covariant derivative regularization supplemented by an arbitrary renormalization prescription that does not break Eq. (\ref{Lambda_Renormalization}) below. After that, using the general statement that the all-loop NSVZ scheme is given by the HD+MSL prescription \cite{Stepanyantz:2020uke}, we obtain the corresponding part of the four-loop gauge $\beta$-function and analyse its scheme dependence. Based on these results, we construct a minimal scheme for the terms under consideration and demonstrate that in this scheme all integrals can be calculated despite the presence of higher derivative regulators. As a correctness test, we verify that the dependence on {\it regularization} parameters disappears (although the renormalization prescription fixes only the {\it renormalization} parameters).

The paper is organized as follows. In Sect. \ref{Section_HD_Regularization} we recall how to regularize ${\cal N}=1$ supersymmetric theories by higher covariant derivatives. The next Sect. \ref{Section_Gamma} is devoted to the calculation of the three-loop contribution to the anomalous dimension of the matter superfields proportional to the sixth power of the Yukawa couplings. In particular, the integrals which determine the one- and two-loop terms of the highest powers in the Yukawa couplings are presented in Subsect. \ref{Subsection_Gamma_Lowest}. After that, the three-loop contribution to the anomalous dimension defined in terms of bare couplings proportional to $(\lambda_0)^6$ is calculated in Subsect. \ref{Subsection_Gamma_Three-Loop}. The corresponding contribution to the standard anomalous dimension (defined in terms of the renormalized couplings) is calculated in Subsect. \ref{Subsection_Gamma_Scheme_Dependence} for an arbitrary renormalization prescription which supplements the higher covariant derivative regularization. The result is also written for the particular cases of the HD+MSL and $\overline{\mbox{DR}}$ schemes. Moreover, in this subsection we construct the minimal scheme and demonstrate that the part of the three-loop anomalous dimension in question in this scheme is independent of both regularization and renormalization parameters. The contribution to the four-loop gauge $\beta$-function proportional to the sixth power of the Yukawa couplings is calculated in Sect. \ref{Section_Beta}. First, we consider the $\beta$-function defined in terms of the bare couplings and obtain such terms with the help of the NSVZ equation. After that, the contribution to the $\beta$-function defined in terms of the renormalized couplings is derived for an arbitrary subtraction scheme supplementing the higher covariant derivative regularization. Analysing the scheme dependence of the result, we construct the expression for this RGF in the minimal scheme. The summary of the results and their discussion are presented in Conclusion. Some technical details of calculations are given in Appendices.

\section{The higher covariant regularization in the supersymmetric case}
\label{Section_HD_Regularization}
\hspace*{\parindent}

In this paper we consider a general renormalizable ${\cal N}=1$ supersymmetric theory with a single gauge coupling in the massless limit. Describing this model it is convenient to use the ${\cal N}=1$ superspace formulation, because in this case supersymmetry becomes a manifest symmetry, see, e.g., \cite{Gates:1983nr,West:1990tg,Buchbinder:1998qv}. At the classical level the superfield action of the theory under consideration can be written as

\begin{equation}\label{Original_Action}
S = \frac{1}{2e_0^2} \mbox{Re}\,\mbox{tr}\int d^4x\,d^2\theta\,W^a W_a + \frac{1}{4}\int d^4x\,d^4\theta\,\phi^{*i} (e^{2V})_i{}^j \phi_j + \Big(\frac{1}{6}\lambda_0^{ijk}\int d^4x\,d^2\theta\,\phi_i\phi_j\phi_k + \mbox{c.c}\Big).
\end{equation}

\noindent
Here $V=V^+$ is the gauge superfield, and $\phi_i$ are chiral matter superfields in a certain representation $R$ of the gauge group $G$. The bare gauge and Yukawa couplings are denoted by $e_0$ and $\lambda_0^{ijk}$, respectively. For the (bare) gauge coupling we will also use the notation $\alpha_0\equiv e_0^2/4\pi$. (In our conventions, the subscript 0 marks the bare couplings.) At the classical level, the supersymmetric gauge field strength is defined as $W_a \equiv \bar D^2 (e^{-2V} D_a e^{2V})/8$ and satisfies the chirality condition $\xbar{D}_{\dot b} W_a = 0$. In the matter part of the action the gauge superfield is written as $V=e_0 V^B T^B$, where $T^B$ are the generators of the representation $R$, while in its gauge part $V= e_0 V^B t^B$, where $t^B$ are the generators of the fundamental representation normalized by the condition $\mbox{tr}(t^A t^B)=\delta^{AB}/2$. The theory is invariant under the gauge transformations

\begin{equation}
\phi\to e^{iA}\phi;\qquad e^{2V}\to e^{iA^+} e^{2V} e^{-iA};\qquad W_a\to e^{iA} W_a  e^{-iA}\equiv \big[e^{iA}\big]_{\mbox{\scriptsize Adj}} W_a
\end{equation}

\noindent
parameterized by a chiral superfield $A = e_0 A^B T^B$ (or $A=e_0 A^B t^B$) if the bare Yukawa couplings satisfy the constraint

\begin{equation}\label{Yukawa_Constraint}
\lambda_0^{mjk} (T^B)_m{}^i + \lambda_0^{imk} (T^B)_m{}^j + \lambda_0^{ijm} (T^B)_m{}^k =0.
\end{equation}

It is convenient to quantize the theory (\ref{Original_Action}) with the help of the background field method, which allows constructing the manifestly gauge invariant effective action. In the superfield formulation it is introduced with the help of the nonlinear background-quantum splitting

\begin{equation}
e^{2V} \to e^{2{\cal F}(v)} e^{2\bm{V}},
\end{equation}

\noindent
where $\bm{V}$ and $v$ denote the background and quantum gauge superfields, respectively. The function ${\cal F}(v) = v+ O(v^3)$ is needed for absorbing the nonlinear terms which appear in the renormalization of the quantum gauge superfield \cite{Piguet:1981fb,Piguet:1981hh,Tyutin:1983rg}. It contains all odd powers of $v$ and in the lowest nontrivial approximation has been derived in \cite{Juer:1982fb,Juer:1982mp}. Subsequently, by an explicit calculation it was demonstrated \cite{Kazantsev:2018kjx} that the nonlinear renormalization is really needed for the renormalization group equations to be satisfied.

Following \cite{Aleshin:2016yvj,Kazantsev:2017fdc}, for introducing the higher covariant derivative regularization we modify the action (\ref{Original_Action}) by adding to it some terms containing higher powers of the covariant derivatives

\begin{equation}\label{Covariant_Derivatives}
\nabla_a \equiv D_a;\qquad \xbar\nabla_{\dot a}\equiv e^{2{\cal F}(v)} e^{2\bm{V}} \xbar D_{\dot a} e^{-2\bm{V}} e^{-2{\cal F}(v)},
\end{equation}

\noindent
where the indices without dots numerate components of the right spinors, and the dotted indices correspond to the left spinors. (Note that the derivatives (\ref{Covariant_Derivatives}) should act on a superfield $X$ which changes as $X\to e^{iA^+} X$ under the gauge transformations.) Then the regularized action can be presented in the form

\begin{eqnarray}\label{Action_Regularized_By_HD}
&& S_{\mbox{\scriptsize reg}} = \frac{1}{2e_0^2} \mbox{Re}\,\mbox{tr}\int d^4x\,d^2\theta\,W^a\bigg[e^{-2\bm{V}} e^{-2{\cal F}(v)} R\Big(-\frac{\xbar\nabla^2\nabla^2}{16\Lambda^2}\Big)e^{2{\cal F}(v)} e^{2\bm{V}}\bigg]_{\mbox{\scriptsize Adj}}W_a\nonumber\\
&& + \frac{1}{4}\int d^4x\,d^4\theta\,\phi^+ F\Big(-\frac{\xbar\nabla^2\nabla^2}{16\Lambda^2}\Big) e^{2{\cal F}(v)} e^{2\bm{V}}\phi + \Big(\frac{1}{6}\lambda_0^{ijk}\int d^4x\,d^2\theta\,\phi_i\phi_j\phi_k + \mbox{c.c}\Big),\qquad
\end{eqnarray}

\noindent
where $\Lambda$ has the dimension of mass. The regulator functions $R$ and $F$ are the sums of 1 coming from the original action (\ref{Original_Action}) and the terms with higher derivatives. In the simplest case it is possible to choose them in the form $R(x)=1+x^n$ and $F(x) = 1+x^m$, where $n$ and $m$ are positive integers. Certainly, these functions can be chosen in an arbitrary way, but should satisfy the conditions $R(0)=F(0)=1$ and rapidly grow at $x\to \infty$. For instance, the exponential functions were used in \cite{Singh:2025fgo}.

After the background-quantum splitting the original gauge invariance generates two different symmetries. The quantum gauge invariance is broken by the gauge fixing procedure, while the background gauge invariance remains a manifest symmetry of the effective action provided that the gauge fixing term does not break it. We will use the background invariant gauge fixing term

\begin{equation}
S_{\mbox{\scriptsize gf}} = - \frac{1}{16 e_0^2\xi_0} \mbox{tr} \int d^4x\,d^4\theta\,\bm{\nabla}^2 v\, R\Big(-\frac{\bm{\xbar\nabla}^2\bm{\nabla}^2}{16\Lambda^2}\Big)_{\mbox{\scriptsize Adj}} \bm{\xbar\nabla}^2 v,
\end{equation}

\noindent
where the background covariant derivatives are given by the expressions

\begin{equation}
\bm{\nabla}_a \equiv D_a;\qquad \bm{\xbar\nabla}_{\dot a} \equiv e^{2\bm{V}} \xbar D_{\dot a} e^{-2\bm{V}}.
\end{equation}

\noindent
The gauge parameter denoted by $\xi_0$ is equal to 1 in the Feynman gauge.

The gauge fixing procedure also requires introducing the Faddeev--Popov and Nielsen--Kallosh ghosts. All details of the corresponding construction can be found in \cite{Aleshin:2016yvj,Kazantsev:2017fdc}. However, in this paper we do not explicitly write down the expressions for their actions because here we will consider only those parts of the effective action that are determined by supergraphs without ghost loops.

We will be interested in the renormalization of the chiral matter superfields, gauge and Yukawa couplings, which is described by the equations

\begin{equation}
\phi_{i} = (\sqrt{Z})_i{}^j\big(\alpha_0,\lambda_0,\ln\Lambda/\mu\big)\, \phi_{j,R};\qquad
\alpha = \alpha\big(\alpha_0,\lambda_0,\ln\Lambda/\mu\big);\qquad
\lambda^{ijk} = \lambda^{ijk}\big(\alpha_0,\lambda_0,\ln\Lambda/\mu\big),
\end{equation}

\noindent
where $\phi_{R,}$, $\alpha$, and $\lambda^{ijk}$ are the renormalized chiral matter superfields, gauge and Yukawa couplings, respectively, and $\mu$ is a renormalization point.

The divergent contributions to the effective action can be conveniently encoded in RGFs. Following \cite{Kataev:2013eta}, we will distinguish various definitions of them. Namely, RGFs defined in terms of the bare couplings by the equations

\begin{eqnarray}\label{RGFs_Bare_Couplings}
&& \gamma_i{}^j(\alpha_0,\lambda_0) \equiv - \frac{d\ln Z_i{}^j}{d\ln\Lambda}\bigg|_{\alpha,\lambda=\mbox{\scriptsize const}};\qquad\
\beta(\alpha_0,\lambda_0) \equiv \frac{d\alpha_0}{d\ln\Lambda}\bigg|_{\alpha,\lambda=\mbox{\scriptsize const}};\nonumber\\
&& (\beta_\lambda)^{ijk}(\alpha_0,\lambda_0) \equiv \frac{d\lambda_0^{ijk}}{d\ln\Lambda}\bigg|_{\alpha,\lambda=\mbox{\scriptsize const}};\qquad
(\beta^*_\lambda)_{ijk}(\alpha_0,\lambda_0) \equiv \frac{d\lambda^*_{0ijk}}{d\ln\Lambda}\bigg|_{\alpha,\lambda=\mbox{\scriptsize const}}
\end{eqnarray}

\noindent
are regularization dependent, but do not depend on a renormalization scheme for a fixed regularization. However, the standard definition is in terms of the renormalized couplings. In our conventions these RGFs are marked by tildes and are introduced by the equations

\begin{eqnarray}\label{RGFs_Renormalized_Couplings}
&& \widetilde\gamma_i{}^j(\alpha,\lambda) \equiv \frac{d\ln Z_i{}^j}{d\ln\mu}\bigg|_{\alpha_0,\lambda_0=\mbox{\scriptsize const}};\qquad\quad
\widetilde\beta(\alpha,\lambda) \equiv \frac{d\alpha}{d\ln\mu}\bigg|_{\alpha_0,\lambda_0=\mbox{\scriptsize const}};\nonumber\\
&& (\widetilde\beta_\lambda)^{ijk}(\alpha,\lambda) \equiv \frac{d\lambda^{ijk}}{d\ln\mu}\bigg|_{\alpha_0,\lambda_0=\mbox{\scriptsize const}};\qquad
(\widetilde\beta^*_\lambda)_{ijk}(\alpha,\lambda) \equiv \frac{d\lambda^*_{ijk}}{d\ln\mu}\bigg|_{\alpha_0,\lambda_0=\mbox{\scriptsize const}}.
\end{eqnarray}

\noindent
They are both regularization and renormalization dependent, and in the HD+MSL scheme reproduce RGFs (\ref{RGFs_Bare_Couplings}) after the formal replacement of arguments $\alpha\to\alpha_0$, $\lambda\to\lambda_0$.

An interesting feature of supersymmetric theories is the existence of certain relations between the $\beta$-functions and the anomalous dimension of the matter superfields, which are satisfied in all orders for certain renormalization prescriptions. Namely, due to the nonrenormalization of the superpotential \cite{Grisaru:1979wc}, the Yukawa couplings can be renormalized as

\begin{equation}\label{Lambda_Renormalization}
\lambda^{ijk} = \lambda_0^{mnp} (\sqrt{Z})_m{}^i (\sqrt{Z})_n{}^j (\sqrt{Z})_p{}^k.
\end{equation}

\noindent
Differentiating this equation with respect to $\ln\Lambda$ at fixed values of the renormalized couplings and using Eq. (\ref{RGFs_Bare_Couplings}) we obtain the exact expressions for Yukawa $\beta$-functions

\begin{eqnarray}\label{Beta_Lambda_Exact}
&& (\beta_\lambda)^{ijk}(\alpha_0,\lambda_0) = \frac{1}{2}\Big(\lambda_0^{ijm} \gamma_m{}^k(\alpha_0,\lambda_0) + \lambda_0^{imk} \gamma_m{}^j(\alpha_0,\lambda_0) + \lambda_0^{mjk} \gamma_m{}^i(\alpha_0,\lambda_0)\Big);\nonumber\\
&& (\beta_\lambda^*)_{ijk}(\alpha_0,\lambda_0) = \frac{1}{2}\Big(\lambda^*_{0ijm} \gamma_k{}^m(\alpha_0,\lambda_0) + \lambda^*_{0imk} \gamma_j{}^m(\alpha_0,\lambda_0) + \lambda^*_{0mjk} \gamma_i{}^m(\alpha_0,\lambda_0)\Big).\qquad
\end{eqnarray}

\noindent
These equations are written for RGFs defined in terms of the bare couplings and in particular are valid for supersymmetric theories regularized by higher derivatives. Similar relations written for RGFs defined in terms of the renormalized couplings are valid in any MS-like scheme and, for instance, in the HD+MSL and $\overline{\mbox{DR}}$ schemes.

In supersymmetric theories the $\beta$-function for the gauge coupling is also related to the anomalous dimension of the matter superfields by the NSVZ equation. For theories with a single gauge coupling it can be written in the form

\begin{equation}\label{Beta_NSVZ}
\beta(\alpha_0,\lambda_0) = -\frac{\alpha_0^2\Big(3C_2-T(R) + C(R)_i{}^j\gamma_j{}^i(\alpha_0,\lambda_0)/r\Big)}{2\pi(1-\alpha_0 C_2/2\pi)},
\end{equation}

\noindent
where $r$ is the dimension of the gauge group $G$ with structure constants $f^{ABC}$,

\begin{equation}
\mbox{tr}(T^A T^B) = T(R)\,\delta^{AB};\qquad (T^A T^A)_i{}^j = C(R)_i{}^j;\qquad f^{ACD} f^{BCD} = C_2\delta^{AB}.
\end{equation}

\noindent
The NSVZ equation can also be generalized to the case of theories with multiple gauge couplings, see \cite{Shifman:1996iy,Korneev:2021zdz}. Eq. (\ref{Beta_NSVZ}) is satisfied in all orders for the RGFs defined in terms of the bare couplings if higher covariant derivatives are used for regularizing a theory \cite{Stepanyantz:2020uke}. Similar relation for RGFs defined in terms of the renormalized couplings is valid for the HD+MSL renormalization prescription. The NSVZ equation (\ref{Beta_NSVZ}) establishes a correspondence between the $\beta$-function in a certain loop and the anomalous dimension of the matter superfields in the previous loops. In the Abelian case
\cite{Vainshtein:1985ynw,Shifman:1985fi} the $\beta$-function is related to the anomalous dimension in the previous loop even at the level of loop integrals \cite{Smilga:2004zr,Kazantsev:2014yna}. In the non-Abelian case \cite{Stepanyantz:2016gtk} the integrals giving the gauge $\beta$-function are related to the integrals giving the anomalous dimensions of the quantum gauge superfield, Faddeev--Popov ghosts, and matter superfields in the previous loop, see, e.g., \cite{Kazantsev:2018nbl,Kuzmichev:2019ywn}. (It is interesting that similar relations appear even for nonrenormalizable theories \cite{Lakhal:2025nbh}.)

According to Eq. (\ref{Beta_NSVZ}), if the anomalous dimension (defined in terms of the bare couplings or in the HD+MSL scheme) is known in a certain loop for a theory regularized by higher covariant derivatives, then the $\beta$-function in the next loop can be obtained immediately. That is why the explicit supergraph calculations in this paper will be made only for the anomalous dimension of the matter superfields. For calculating the anomalous dimension, we first consider a part of the effective action corresponding to the two-point Green function of the matter superfield. It can be presented in the form

\begin{equation}\label{G_Function_Definition}
\Gamma^{(2)}_\phi = \frac{1}{4} \int \frac{d^4p}{(2\pi)^4}\,d^4\theta\,\phi^{*i}(-p,\theta)\,\phi_j(p,\theta)\,G_i{}^j\big(\alpha_0,\lambda_0,\ln\Lambda/p\big),
\end{equation}

\noindent
where in the tree approximation the function $G_i{}^j$ is equal to $\delta_i^j$. Taking into account that the renormalized function $ZG$ is finite in the limit $\Lambda\to \infty$, it is possible to present the anomalous dimension defined in terms of the bare couplings in the form

\begin{eqnarray}\label{Lambda_Equation}
&&\hspace*{-7mm} \gamma_i{}^j(\alpha_0,\lambda_0) = \frac{d}{d\ln\Lambda} \ln G_i{}^j\big(\alpha_0,\lambda_0,\ln\frac{\Lambda}{p}\big)\bigg|_{\alpha,\lambda=\mbox{\scriptsize const};\,\Lambda\to \infty} = \lim\limits_{p\to 0}\bigg(\frac{\partial\ln G_i{}^j}{\partial\ln\Lambda} + \frac{\partial\ln G_i{}^j}{\partial\alpha_0}\, \beta(\alpha_0,\lambda_0)\nonumber\\
&&\hspace*{-7mm} + \frac{\partial\ln G_i{}^j}{\partial\lambda_0^{mnp}}\, (\beta_\lambda)^{mnp}(\alpha_0,\lambda_0) +\frac{\partial\ln G_i{}^j}{\partial\lambda^*_{0mnp}}\, (\beta^*_{\lambda})_{mnp}(\alpha_0,\lambda_0) \bigg).
\end{eqnarray}

\noindent
Below this equation will be used for calculating the anomalous dimension because it allows to do this without involving the relations between the bare and renormalized couplings, as well as the explicit form of the renormalization constant for the matter superfields.

\section{The three-loop anomalous dimension of the matter superfields}
\label{Section_Gamma}

\subsection{The highest-order Yukawa terms in the one- and two-loop approximations}
\label{Subsection_Gamma_Lowest}
\hspace*{\parindent}

\begin{figure}[h]
\begin{picture}(0,2)
\put(4.0,0){\includegraphics[scale=0.18]{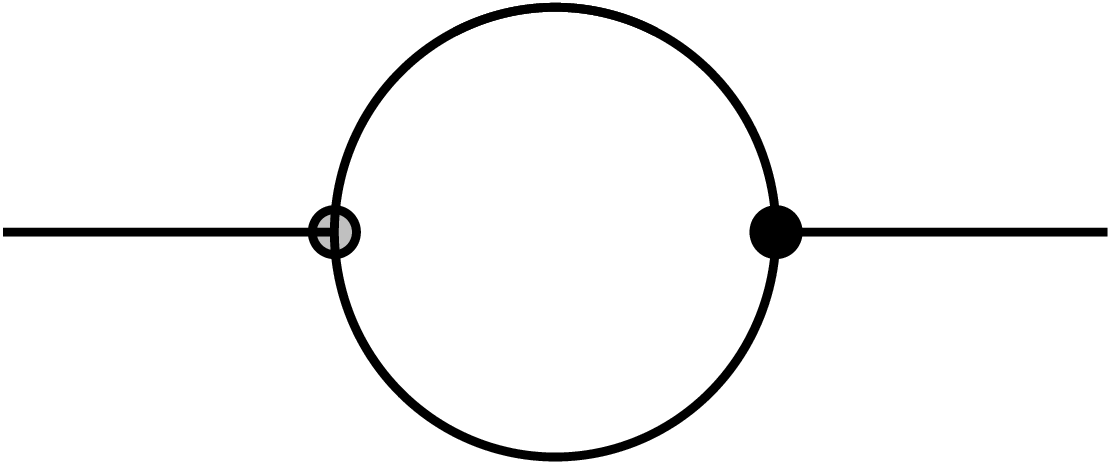}}
\put(4.1,1.2){(1)}
\put(8.2,0){\includegraphics[scale=0.18]{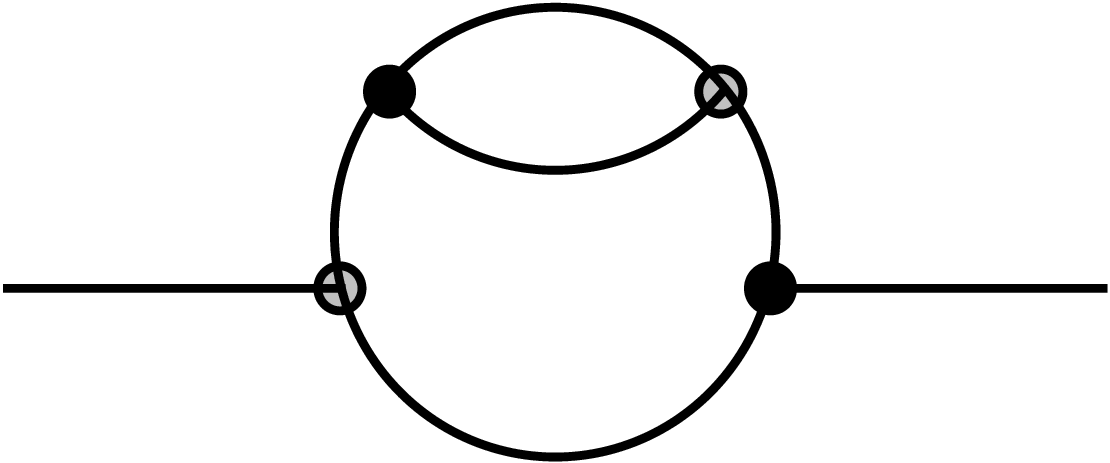}}
\put(8.3,1.2){(2)}
\end{picture}
\caption{The one- and two-loop superdiagrams contributing to the anomalous dimension of the matter superfields which contain only Yukawa vertices}\label{Figure_Lowest_Loop}
\end{figure}

Let us first calculate the three-loop contribution proportional to the sixth power of the Yukawa couplings to the anomalous dimension of the matter superfields defined in terms of the bare couplings. Note that, for this purpose, we also need to find the leading (in the Yukawa couplings) terms in the one- and two-loop approximations. They are given by the supergraphs presented in Fig. \ref{Figure_Lowest_Loop}. (Note that the supergraphs containing internal gauge lines are not included because they are proportional to the gauge coupling constant and are not essential for obtaining the part of the anomalous dimension considered in this paper. Moreover, we omitted a two-loop supergraph giving vanishing contribution in the massless limit.) In our conventions, the filled disks denote the triple vertices proportional to $\lambda_0^{ijk}$, while the circles correspond to the $\lambda^*_{0ijk}$ vertices.

In the case of using the higher covariant derivative regularization all these supergraphs have already been calculated in \cite{Shakhmanov:2017soc,Kazantsev:2018nbl} (see also \cite{Kazantsev:2020kfl}). Here we briefly repeat this calculation by a slightly different method for illustrating the technique used in what follows for obtaining a more complicated three-loop contribution to the anomalous dimension. In terms of the loop integrals, the two-loop anomalous dimension is given by the expression

\begin{eqnarray}\label{One-Loop_Gamma}
&&\hspace*{-6mm} \gamma_i{}^j(\alpha_0,\lambda_0) = \frac{d\ln G_i{}^j}{d\ln\Lambda}\bigg|_{p\to 0} = \frac{d}{d\ln\Lambda} \bigg\{\lambda^*_{0iab} \lambda_0^{jab} \int \frac{d^4K}{(2\pi)^4} \frac{2}{K^4 F_K^2}
- (\lambda^*_{0iab}\lambda_0^{kab}) (\lambda^*_{0kcd}\lambda_0^{jcd}) \nonumber\\
&&\hspace*{-6mm} \times \int \frac{d^4K}{(2\pi)^4}  \frac{d^4L}{(2\pi)^4} \frac{2}{K^4 F_K^2 L^4 F_L^2}
- \lambda^*_{0iab} \lambda_0^{jac} \lambda^*_{0cde} \lambda_0^{bde} \int \frac{d^4K}{(2\pi)^4}  \frac{d^4L}{(2\pi)^4} \frac{8}{K^4 F_K^3 L^2 F_L (K+L)^2 F_{K+L}}
\nonumber\\
&&\hspace*{-6mm} + O(\alpha_0)+\mbox{higher orders}\bigg\},
\end{eqnarray}

\noindent
where $O(\alpha_0)$ denotes terms proportional to positive powers of $\alpha_0$ and, therefore, includes all one- and two-loop terms coming from superdiagrams containing internal gauge lines. In our notation, Euclidean momenta are denoted by capital letters and $F_K \equiv F(K^2/\Lambda^2)$. Note that the derivative with respect to $\ln\Lambda$ can act on the bare couplings $\alpha_0$ and $\lambda_0^{ijk}$. The corresponding contributions have larger powers of couplings and should be taken into account in {\it all} subsequent approximations.

In the one-loop approximation the loop integral can easily be taken for a general regulator function,

\begin{equation}\label{One-Loop_Integral}
\int \frac{d^4K}{(2\pi)^4} \frac{d}{d\ln\Lambda} \frac{1}{K^4 F_K^2} = - 2 \int \frac{d^4K}{(2\pi)^4} \frac{1}{K^4} \frac{d}{d\ln K^2} \frac{1}{F_K^2}
=  -\frac{1}{8\pi^2} \frac{1}{F_K^2}\bigg|_0^\infty = \frac{1}{8\pi^2},\qquad
\end{equation}

\noindent
so that the one-loop anomalous dimension (defined in terms of the bare couplings) takes the form

\begin{eqnarray}\label{Gamma_1Loop}
&& \gamma_i{}^j(\alpha_0,\lambda_0) = \lambda^*_{0iab} \lambda_0^{jab} \int \frac{d^4K}{(2\pi)^4}  \frac{d}{d\ln\Lambda} \frac{2}{K^4 F_K^2} + O(\alpha_0)+\mbox{higher orders}\nonumber\\
&&\qquad\qquad\qquad\qquad\qquad\qquad\qquad\qquad\qquad = \frac{1}{4\pi^2}\lambda^*_{0iab} \lambda_0^{jab} + O(\alpha_0)+\mbox{higher orders}.\qquad\quad
\end{eqnarray}

In the next two-loop approximation it is necessary to take into account the terms obtained by differentiating $\lambda_0$ with respect to $\ln\Lambda$,

\begin{eqnarray}\label{Gamma_2Loop_Original_Integrals}
&&\hspace*{-5mm} \gamma_i{}^j(\alpha_0,\lambda_0) = \frac{1}{4\pi^2}\lambda^*_{0iab} \lambda_0^{jab}
+ \frac{d}{d\ln\Lambda} \Big(\lambda^*_{0iab} \lambda_0^{jab}\Big) \cdot \int \frac{d^4K}{(2\pi)^4} \frac{2}{K^4 F_K^2}
- (\lambda^*_{0iab}\lambda_0^{kab}) (\lambda^*_{0kcd}\lambda_0^{jcd}) \nonumber\\
&&\hspace*{-5mm} \times \int \frac{d^4K}{(2\pi)^4}  \frac{d^4L}{(2\pi)^4} \frac{d}{d\ln\Lambda} \frac{2}{K^4 F_K^2 L^4 F_L^2}
- \lambda^*_{0iab} \lambda_0^{jac} \lambda^*_{0cde} \lambda_0^{bde} \int \frac{d^4K}{(2\pi)^4}  \frac{d^4L}{(2\pi)^4} \frac{d}{d\ln\Lambda} \frac{8}{K^4 F_K^3 L^2 F_L}\nonumber\\
&&\hspace*{-5mm} \times \frac{1}{(K+L)^2 F_{K+L}} + O(\alpha_0)+\mbox{higher orders}.
\end{eqnarray}

\noindent
The derivatives of the Yukawa couplings are calculated with the help of Eqs. (\ref{RGFs_Bare_Couplings}) and (\ref{Beta_Lambda_Exact}). The result is expressed in terms of the anomalous dimension of the matter superfields. When calculating the contribution in question, it is necessary to take only its one-loop part quadratic in Yukawa couplings given by Eq. (\ref{Gamma_1Loop}). Moreover, instead of using the final expression, it is more convenient to present this part of the anomalous dimension in the form of a loop integral,

\begin{eqnarray}
&&\hspace*{-5mm} \frac{d}{d\ln\Lambda} \Big(\lambda^*_{0iab} \lambda_0^{jab}\Big) = \lambda^*_{0iab} \Big(\lambda_0^{kab} \gamma_k{}^j(\alpha_0,\lambda_0) + 2 \lambda_0^{jcb}\gamma_c{}^a(\alpha_0\lambda_0)\Big)
=  (\lambda^*_{0iab}\lambda_0^{kab}) (\lambda^*_{0kcd}\lambda_0^{jcd})
\nonumber\\
&&\hspace*{-5mm} \times \int \frac{d^4L}{(2\pi)^4}  \frac{d}{d\ln\Lambda} \frac{2}{L^4 F_L^2} + \lambda^*_{0iab} \lambda_0^{jac} \lambda^*_{0cde} \lambda_0^{bde} \int \frac{d^4L}{(2\pi)^4}  \frac{d}{d\ln\Lambda} \frac{4}{L^4 F_L^2} + O(\alpha_0) + \mbox{higher orders}.\nonumber\\
\end{eqnarray}

Substituting this expression into Eq. (\ref{Gamma_2Loop_Original_Integrals}) we present the two-loop contribution to the anomalous dimension as the {\it well-defined} loop integral

\begin{eqnarray}\label{Gamma_2Loop_Integrals}
&&\hspace*{-5mm} \gamma_i{}^j(\alpha_0,\lambda_0) = \frac{1}{4\pi^2}\lambda^*_{0iab} \lambda_0^{jab}
- \lambda^*_{0iab} \lambda_0^{jac} \lambda^*_{0cde} \lambda_0^{bde} \int \frac{d^4K}{(2\pi)^4}  \frac{d^4L}{(2\pi)^4} \frac{d}{d\ln\Lambda} \Big(\frac{8}{K^4 F_K^3 L^2 F_L (K+L)^2}\nonumber\\
&&\hspace*{-5mm} \times \frac{1}{F_{K+L}} - \frac{4}{K^4 F_K^2 L^4 F_L^2}\Big) + O(\alpha_0) +\mbox{higher orders}.
\end{eqnarray}

\noindent
Although this integral contains the undefined function $F$, it can nevertheless be calculated. The details of this calculation made with the help of the technique based on the Chebyshev polynomials \cite{Rosner:1967zz}
are presented in Appendix \ref{Appendix_Two-Loop}. The result appears to be independent of a particular choice of the regulator function $F$ and is given by the expression\footnote{With the higher covariant derivative regularization it was first obtained in \cite{Shakhmanov:2017soc} by a different method.}

\begin{eqnarray}\label{Gamma_2Loop_Result}
&&\hspace*{-5mm} \gamma_i{}^j(\alpha_0,\lambda_0) = \frac{1}{4\pi^2}\lambda^*_{0iab} \lambda_0^{jab}
- \frac{1}{16\pi^4} \lambda^*_{0iab} \lambda_0^{jac} \lambda^*_{0cde} \lambda_0^{bde} + O(\alpha_0) +\mbox{higher orders}.
\end{eqnarray}

\noindent
This expression gives (a part of) the anomalous dimension defined in terms of the bare couplings. The anomalous dimension standardly defined in terms of the renormalized couplings will be obtained in what follows after calculating the three-loop terms proportional to the sixth power of Yukawa couplings.

\subsection{The three-loop contributions proportional to the sixth power of the Yukawa couplings}
\label{Subsection_Gamma_Three-Loop}
\hspace*{\parindent}

\begin{figure}[h]
\begin{picture}(0,3)
\put(0.3,0.31){\includegraphics[scale=0.18]{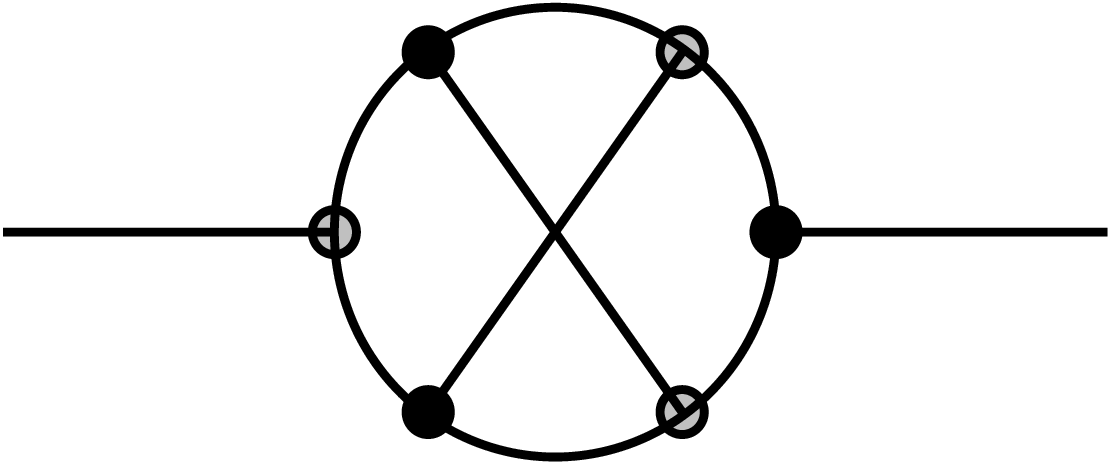}}
\put(0.5,1.8){(1)}
\put(4.3,0){\includegraphics[scale=0.18]{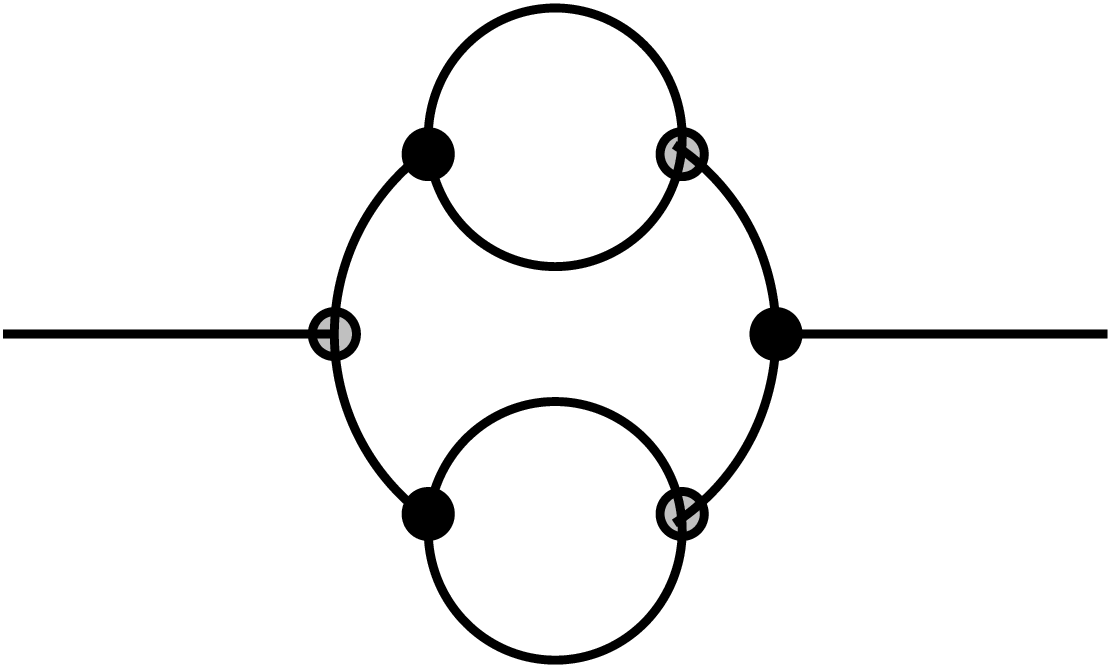}}
\put(4.5,1.8){(2)}
\put(8.3,0.29){\includegraphics[scale=0.18]{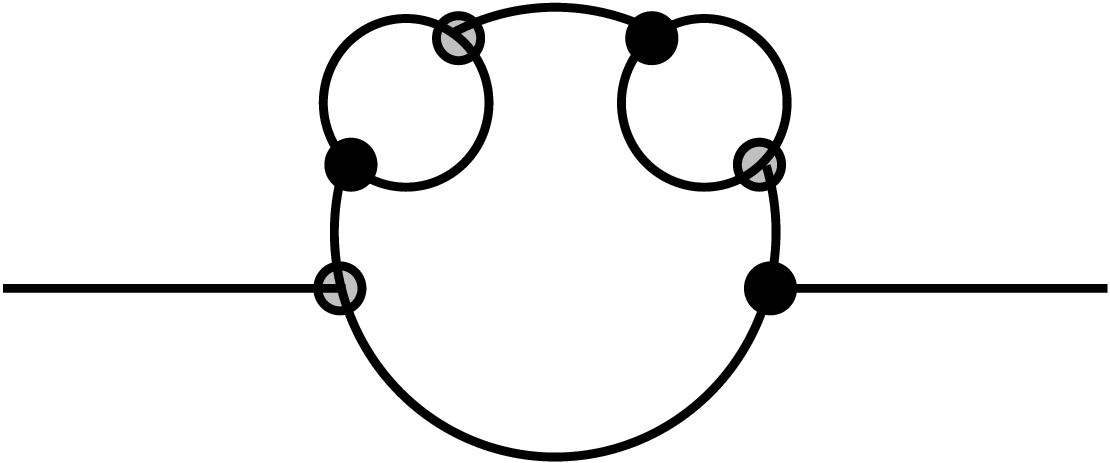}}
\put(8.5,1.8){(3)}
\put(12.3,0.29){\includegraphics[scale=0.18]{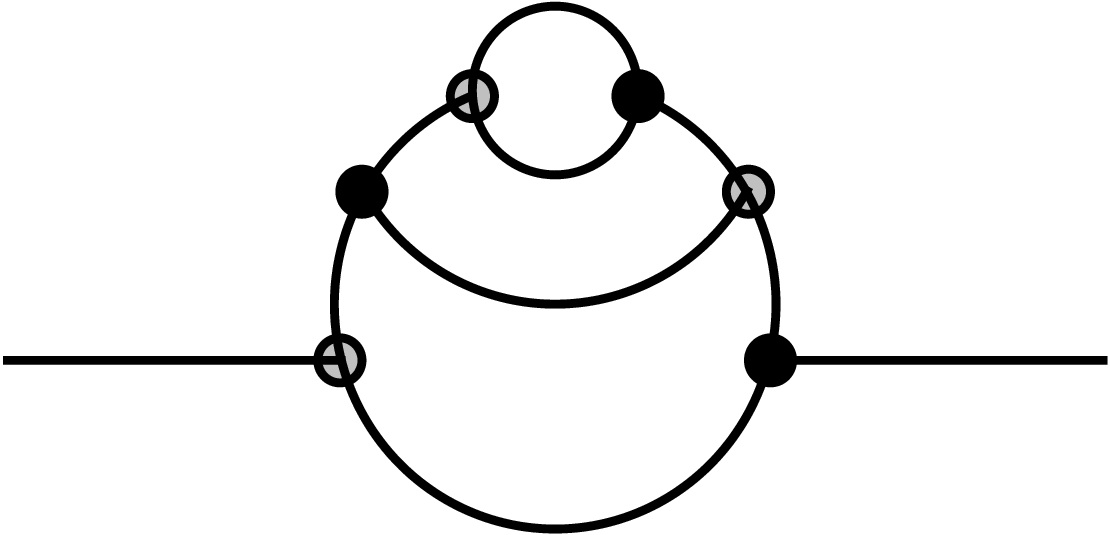}}
\put(12.5,1.8){(4)}
\end{picture}
\caption{The three-loop superdiagrams proportional to the sixth power of the Yukawa couplings contributing to the anomalous dimension of the matter superfields}\label{Figure_Lambda6}
\end{figure}
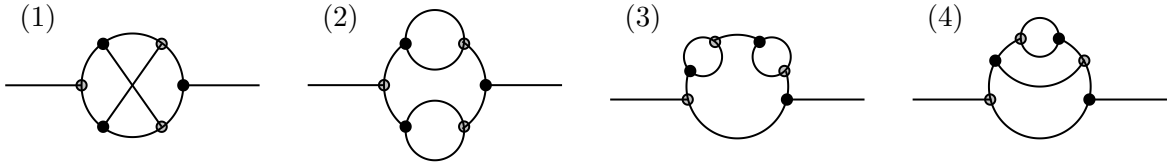

The supergraphs which determine the two-point Green function of the matter superfields proportional to the sixth power of the Yukawa couplings in the three-loop approximation are presented in Fig. \ref{Figure_Lambda6}.

To obtain the corresponding contribution to the anomalous dimension, we use the same technique that was described in the previous subsection. Namely, we find the expressions for all these diagrams in the form of loop integrals. The results are collected in Appendix \ref{Appendix_Diagrams}. Next, the expression for the part of the function $G$ under consideration is constructed by summing one-, two- and three-loop contributions written in the form of loop integrals. After that, we calculate $\ln G$ (again in terms of loop integrals) and differentiate the result with respect to $\ln\Lambda$ at fixed values of the renormalized couplings. Exactly as in the previous subsection, the derivatives of the Yukawa couplings present in the one- and two-loop terms contribute to the three-loop expression for the anomalous dimension. After some simple, but rather lengthy calculations we obtained that the anomalous dimension can be presented in the form

\begin{equation}
\gamma_i{}^j(\alpha_0,\lambda_0) = \frac{1}{4\pi^2}\lambda^*_{0iab} \lambda_0^{jab}
- \frac{1}{16\pi^4} \lambda^*_{0iab} \lambda_0^{jac} \lambda^*_{0cde} \lambda_0^{bde} + \Delta_3\gamma_i{}^j + O(\alpha_0) + \mbox{higher orders},
\end{equation}

\noindent
where the three-loop contribution is given by the expression

\begin{eqnarray}\label{Gamma_3Loop}
&& \Delta_3\gamma_i{}^j = \lambda_0^{jkl} \lambda^*_{0lrs} \lambda_0^{pmr} \lambda^*_{0kmn} \lambda_0^{qns} \lambda^*_{0ipq}\, I_1
+ \lambda_0^{jml} (\lambda^*_{0mkp} \lambda_0^{pab} \lambda^*_{0qab} \lambda_0^{nkq}) \lambda^*_{0inl}\, I_2 \nonumber\\
&&\qquad\quad + \Big(2\lambda_0^{jmp} (\lambda^*_{0mab} \lambda_0^{nab})(\lambda^*_{0ncd}\lambda_0^{kcd}) \lambda^*_{0ikp}
+ \lambda_0^{jmn} (\lambda^*_{0mab} \lambda_0^{pab}) (\lambda^*_{0ncd} \lambda_0^{qcd}) \lambda^*_{0ipq}\Big)\,I_3,\qquad
\end{eqnarray}

\noindent
with $I_i$ being the following loop integrals:

\begin{eqnarray}\label{Integral_I1}
&&\hspace*{-5mm} I_1 \equiv \int \frac{d^4Q}{(2\pi)^4}\,\frac{d^4K}{(2\pi)^4}\, \frac{d^4L}{(2\pi)^4}\,
\frac{d}{d\ln\Lambda}\,\frac{16}{K^2 F_K^2 (Q+K+L)^2 F_{Q+K+L}^2}\nonumber\\
&&\hspace*{-5mm}\qquad\qquad\qquad\qquad\qquad\qquad \times \frac{1}{L^2 F_L Q^2 F_Q (K+L)^2 F_{K+L} (K+Q)^2 F_{K+Q}} = \frac{3\zeta(3)}{64\pi^6};\qquad\\
\label{Integral_I2}
&&\hspace*{-5mm} I_2 \equiv 32 \int \frac{d^4Q}{(2\pi)^4}\, \frac{d^4K}{(2\pi)^4}\, \frac{d^4L}{(2\pi)^4}\, \frac{d}{d\ln\Lambda}\bigg\{
\frac{1}{K^4 F_K^3 L^2 F_L^2 (K+L)^2 F_{K+L} Q^2 F_Q (Q+L)^2 F_{Q+L}}\nonumber\\
&&\hspace*{-5mm}\qquad - \frac{1}{Q^4 F_Q^2 K^4 F_K^3 L^2 F_L (K+L)^2 F_{K+L}} + \frac{1}{3 Q^4 F_Q^2 K^4 F_K^2 L^4 F_L^2} \bigg\} = \frac{1}{32\pi^6};\\
\label{Integral_I3}
&&\hspace*{-5mm} I_3 \equiv \int\frac{d^4Q}{(2\pi)^4}\, \frac{d^4K}{(2\pi)^4}\, \frac{d^4L}{(2\pi)^4}\, \bigg\{ \frac{d}{d\ln\Lambda}\frac{8}{K^4F_K^4 L^2 F_L Q^2 F_Q (K+L)^2 F_{K+L} (K+Q)^2 F_{K+Q}}\nonumber\\
&&\hspace*{-5mm}\qquad - \frac{16}{K^4 F_K^3 L^2 F_L (K+L)^2 F_{K+L}}\cdot \frac{d}{d\ln\Lambda} \frac{1}{Q^4 F_Q^2}\bigg\} \equiv \frac{C}{64\pi^6}.
\end{eqnarray}

The integral $I_1$ has been calculated in \cite{Kuzmichev:2023zxy} with the help of the Chebyshev polynomials. The result for it does not depend on a particular form of the regulator function $F(x)$. The integral $I_2$ is also regularization independent. Its calculation is described in Appendix \ref{Appendix_Three-Loop}. However, the integral $I_3$ depends on a particular form of the function $F(x)$ and cannot be calculated if this function is not specified. Nevertheless, this integral is a finite (regularization dependent) constant. It is convenient to take out the multiplier $1/64\pi^6$ introducing the constant $C\equiv 64\pi^6 I_3$.

Thus, the part of the three-loop anomalous dimension of the matter superfields containing terms with the largest powers of the Yukawa couplings takes the form

\begin{eqnarray}\label{Gamma_Bare_Result}
&&\hspace*{-5mm} \gamma_i{}^j(\alpha_0,\lambda_0) = \frac{1}{4\pi^2}\lambda^*_{0iab} \lambda_0^{jab} - \frac{1}{16\pi^4} \lambda^*_{0iab} \lambda_0^{jac} \lambda^*_{0cde} \lambda_0^{bde}
+ \frac{3\zeta(3)}{64\pi^6}\,\lambda_0^{jkl} \lambda^*_{0lrs} \lambda_0^{pmr} \lambda^*_{0kmn} \lambda_0^{qns} \lambda^*_{0ipq}
\nonumber\\
&&\hspace*{-5mm} + \frac{1}{32\pi^6}\,\lambda_0^{jml} (\lambda^*_{0mkp} \lambda_0^{pab} \lambda^*_{0qab} \lambda_0^{nkq}) \lambda^*_{0inl} + \frac{C}{64\pi^6} \Big(2\lambda_0^{jmp} (\lambda^*_{0mab} \lambda_0^{nab})(\lambda^*_{0ncd}\lambda_0^{kcd}) \lambda^*_{0ikp} + \lambda_0^{jmn} \nonumber\\
&&\hspace*{-5mm} \times (\lambda^*_{0mab} \lambda_0^{pab}) (\lambda^*_{0ncd} \lambda_0^{qcd}) \lambda^*_{0ipq}\Big) + O(\alpha_0) + \mbox{higher orders}.\vphantom{\frac{1}{2}}
\end{eqnarray}

\noindent
(Note that all terms proportional to the gauge coupling constant in this expression are not written explicitly and are included into $O(\alpha_0)$.) According to Appendix \ref{Appendix_Three-Loop}, the regularization dependent finite constant $C$ in Eq. (\ref{Gamma_Bare_Result}) is given by

\begin{equation}\label{C_Result}
C = - \frac{1}{4}\int\limits_0^\infty dx \Big(\ln x+ B-1\Big)^2 \frac{d}{dx}\Big(\frac{1}{F^3(x)}\Big) - \int\limits_0^\infty \frac{dx\,f(x)}{x F^3(x)},
\end{equation}

\noindent
where the constant $B$ is defined by Eq. (\ref{B_Definition}) and the function $f(x)$ is determined by Eq. (\ref{Auxiliary_Integral}). As any RGF defined in terms of the bare couplings, the expression (\ref{Gamma_Bare_Result}) is regularization dependent, but does not depend on a particular way of removing divergences for a fixed regularization.

\subsection{The anomalous dimension defined in terms of the renormalized couplings}
\label{Subsection_Gamma_Scheme_Dependence}
\hspace*{\parindent}

Let us calculate the anomalous dimension of the matter superfields standardly defined in terms of the renormalized couplings. Certainly, as earlier, we will consider only the part containing leading terms in the Yukawa couplings. For this purpose, we first substitute the anomalous dimension (\ref{Gamma_Bare_Result}) into the renormalization group equation for $\ln Z$ in Eq. (\ref{RGFs_Bare_Couplings}) and integrate it at fixed values of the renormalized couplings. The result will contain some finite constants which specify the renormalization prescription in the considered order of perturbation theory. In the two-loop approximation the terms containing only Yukawa couplings can be written in the form

\begin{eqnarray}\label{Ln(Z)}
&&\hspace*{-7mm} (\ln Z)_i{}^j = -\frac{1}{4\pi^2} \lambda^*_{imn} \lambda^{jmn} \Big( \ln\frac{\Lambda}{\mu} +g_{12} \Big)
-\frac{1}{8\pi^4} \lambda^*_{iac} \lambda^{jab} \lambda^*_{bde} \lambda^{cde} \Big(\frac{1}{2} \ln^2\frac{\Lambda}{\mu} +g_{12} \ln\frac{\Lambda}{\mu} -\frac{1}{2} \ln\frac{\Lambda}{\mu} \nonumber\\
&&\hspace*{-7mm} +g_{23}\Big) -\frac{1}{16\pi^4} \lambda^*_{iab} \lambda^{kab} \lambda^*_{kcd} \lambda^{jcd} \Big( \frac{1}{2} \ln^2\frac{\Lambda}{\mu} +g_{12} \ln\frac{\Lambda}{\mu} +g_{24}\Big) + O(\alpha) + \mbox{higher orders},\qquad
\end{eqnarray}

\noindent
where we follow the notations adopted in \cite{Kazantsev:2018nbl} and include all terms proportional to the gauge coupling into $O(\alpha)$. Note that for calculating the three-loop anomalous dimension we need the expression for $\ln Z$ in the next, three-loop approximation. It is rather large and is not presented here. However, it was calculated, and, as a correctness test, we have verified that all coefficients at higher logarithms in the HD+MSL scheme agree with the general expressions following from the renormalization group analysis \cite{Derkachev:2017nhd,Meshcheriakov:2022tyi,Meshcheriakov:2023fmk,Meshcheriakov:2024qwj,Kovyrshin:2025ufp}.\footnote{More precisely, we have verified that Eq. (33) in \cite{Kovyrshin:2025ufp} produces the correct result for $(\ln Z)_i{}^j$ in the HD+MSL scheme.} Using Eqs. (\ref{Lambda_Renormalization}) and (\ref{Ln(Z)}) one can construct the equation relating the bare and renormalized Yukawa couplings, which is rather lengthy and is not also presented here. Nevertheless, it should be stressed that everywhere in this paper we use only renormalization prescriptions for which the renormalization of the Yukawa couplings is related to the renormalization of the chiral matter superfields by Eq. (\ref{Lambda_Renormalization}). (In principle, it is possible to consider more general renormalization prescriptions for which this equation is broken, see, e.g., \cite{Jack:1996qq,Haneychuk:2022qvu}.) The finite constants $g_{12}$, $g_{23}$, and $g_{24}$ in Eq. (\ref{Ln(Z)}) (partially) specify the renormalization prescription. The anomalous dimension of the matter superfields depends on them starting from the two-loop order.

After calculating $\ln Z$ in the three-loop approximation\footnote{Note that Eq. (\ref{Ln(Z)}) is written in the {\it two-loop} order.} it is necessary to express the renormalized Yukawa couplings in terms of the bare ones and differentiate the result with respect to $\ln\mu$ according to Eq. (\ref{RGFs_Renormalized_Couplings}). After that, the bare couplings should be inversely expressed in terms of renormalized ones. The result obtained after these transformations is given by the expression

\begin{eqnarray}\label{Gamma_Renormalized_Result}
&&\hspace*{-5mm} \widetilde\gamma_i{}^j(\alpha,\lambda) = \frac{1}{4\pi^2}\lambda^*_{iab} \lambda^{jab} - \frac{1}{16\pi^4} \lambda^*_{iab} \lambda^{jac} \lambda^*_{cde} \lambda^{bde}
+ \frac{3\zeta(3)}{64\pi^6}\,\lambda^{jkl} \lambda^*_{lrs} \lambda^{pmr} \lambda^*_{kmn} \lambda^{qns} \lambda^*_{ipq} + \frac{1}{32\pi^6}
\nonumber\\
&&\hspace*{-5mm} \times \lambda^{jml} (\lambda^*_{mkp} \lambda^{pab} \lambda^*_{qab} \lambda^{nkq}) \lambda^*_{inl} + \frac{1}{64\pi^6} \lambda^{jmp} (\lambda^*_{mab} \lambda^{nab})(\lambda^*_{ncd}\lambda^{kcd}) \lambda^*_{ikp}\Big(2C + g_{12}^2 - 2g_{12} - 4g_{23} \nonumber\\
&&\hspace*{-5mm} + 2g_{24} \Big) + \frac{1}{64\pi^6} \lambda^{jmn} (\lambda^*_{mab} \lambda^{pab}) (\lambda^*_{ncd} \lambda^{qcd}) \lambda^*_{ipq}\Big(C + g_{12}^2 - g_{12} - 2g_{23}\Big)
+ \frac{1}{128\pi^6}\Big((\lambda^*_{iab}\lambda^{kab})
\nonumber\\
&&\hspace*{-5mm} \times(\lambda^*_{kcd}\lambda^{lcd})(\lambda^*_{lef}\lambda^{jef}) + 4(\lambda^*_{iab}\lambda^{kab})(\lambda^*_{kcd} \lambda^{jce} \lambda^*_{efg} \lambda^{dfg})\Big)\big(g_{12}^2 - 2g_{24}\big) + O(\alpha) + \mbox{higher orders}.\vphantom{\frac{1}{2}}\nonumber\\
\end{eqnarray}

\noindent
(Again, here we do not explicitly write down the terms proportional to the gauge coupling constant $\alpha$. In the case of using the higher covariant derivative regularization, the one- and two-loop expressions for them can be found in \cite{Kazantsev:2020kfl}.)

It is expedient to single out three particular renormalization prescriptions.

\medskip

\textbf{1)} The \textbf{HD+MSL scheme} is obtained if all finite constants are set to 0,

\begin{equation}\label{Gamma_Constants_HD+MSL}
g_{12} = 0;\qquad g_{23} = 0;\qquad g_{24} = 0.
\end{equation}

\noindent
In this case (in agreement with the general statement \cite{Kataev:2013eta,Stepanyantz:2020uke}) the anomalous dimension (\ref{Gamma_Renormalized_Result}) coincides with the one defined in terms of the bare couplings after the formal replacement of arguments $\alpha_0\to\alpha$, $\lambda_0^{ijk}\to \lambda^{ijk}$,

\begin{eqnarray}\label{Gamma_HD+MSL}
&&\hspace*{-5mm} \widetilde\gamma_i{}^j(\alpha,\lambda)\Big|_{\mbox{\scriptsize HD+MSL}} = \frac{1}{4\pi^2}\lambda^*_{iab} \lambda^{jab} - \frac{1}{16\pi^4} \lambda^*_{iab} \lambda^{jac} \lambda^*_{cde} \lambda^{bde}
+ \frac{3\zeta(3)}{64\pi^6}\,\lambda^{jkl} \lambda^*_{lrs} \lambda^{pmr} \lambda^*_{kmn} \lambda^{qns} \lambda^*_{ipq}
\nonumber\\
&&\hspace*{-5mm} + \frac{1}{32\pi^6}\,\lambda^{jml} (\lambda^*_{mkp} \lambda^{pab} \lambda^*_{qab} \lambda^{nkq}) \lambda^*_{inl} + \frac{C}{64\pi^6} \Big(2\lambda^{jmp} (\lambda^*_{mab} \lambda^{nab})(\lambda^*_{ncd}\lambda^{kcd}) \lambda^*_{ikp} + \lambda^{jmn} (\lambda^*_{mab} \lambda^{pab})\nonumber\\
&&\hspace*{-5mm} \times (\lambda^*_{ncd} \lambda^{qcd}) \lambda^*_{ipq}\Big) + O(\alpha) + \mbox{higher orders}.\vphantom{\frac{1}{2}}
\end{eqnarray}

\noindent
Although for this renormalization prescription the NSVZ equation is valid in all orders, RGFs in this scheme depend on the regularization parameters. In particular, the expression (\ref{Gamma_HD+MSL}) contains the constant $C$ given by Eq. (\ref{C_Result}), which, in turn, depends on a particular choice of the regulator function $F$ in Eq. (\ref{Action_Regularized_By_HD}).

\medskip

\textbf{2)} For the $\overline{\mbox{\textbf{DR}}}$ \textbf{scheme} the result for the three-loop anomalous dimension can be found in \cite{Jack:1996qq}. In our notation the part containing only leading terms in the Yukawa couplings takes the form

\begin{eqnarray}\label{Gamma_DR_Bar}
&&\hspace*{-5mm} \widetilde\gamma_i{}^j(\alpha,\lambda)\Big|_{\overline{\mbox{\scriptsize DR}}} = \frac{1}{4\pi^2}\lambda^*_{iab} \lambda^{jab} - \frac{1}{16\pi^4} \lambda^*_{iab} \lambda^{jac} \lambda^*_{cde} \lambda^{bde}
+ \frac{3\zeta(3)}{64\pi^6}\,\lambda^{jkl} \lambda^*_{lrs} \lambda^{pmr} \lambda^*_{kmn} \lambda^{qns} \lambda^*_{ipq}
\nonumber\\
&&\hspace*{-5mm} + \frac{1}{32\pi^6} \lambda^{jml} (\lambda^*_{mkp} \lambda^{pab} \lambda^*_{qab} \lambda^{nkq}) \lambda^*_{inl} - \frac{1}{128\pi^6} \lambda^{jmp} (\lambda^*_{mab} \lambda^{nab})(\lambda^*_{ncd}\lambda^{kcd}) \lambda^*_{ikp}  - \frac{1}{256\pi^6} \lambda^{jmn} \nonumber\\
&&\hspace*{-5mm} \times (\lambda^*_{mab} \lambda^{pab})(\lambda^*_{ncd} \lambda^{qcd}) \lambda^*_{ipq}
+ O(\alpha) + \mbox{higher orders}.\vphantom{\frac{1}{2}}
\end{eqnarray}

\noindent
For completeness, here we recall the relation between the notations adopted in this paper and the ones in Refs. \cite{Jack:1996vg,Jack:1996cn,Jack:1996qq,Jack:1998uj},

\begin{equation}\label{Translation}
\alpha = \frac{g^2}{4\pi};\qquad \lambda^{ijk} = \frac{1}{2} Y^{ijk};\qquad \widetilde{\gamma}_i{}^j(\alpha,\lambda)\Big|_{\overline{\mbox{\scriptsize DR}}} = 2\gamma^j{}_i(g, Y);\qquad
\widetilde\beta(\alpha,\lambda)\Big|_{\overline{\mbox{\scriptsize DR}}} = \frac{g}{2\pi} \beta(g,Y).
\end{equation}

\noindent
The expression (\ref{Gamma_DR_Bar}) is rather simple and does not contain arbitrary parameters. However, the NSVZ equation does not hold for the $\overline{\mbox{DR}}$ scheme. This is one of its main drawbacks (alongside the use  of the space-time with non-integer dimension).

Certainly, the result for the $\overline{\mbox{DR}}$ renormalization prescription should be obtained from the expression (\ref{Gamma_Renormalized_Result}). According to \cite{Kazantsev:2020kfl}, the value of the constant $g_{12}$ which allows reproducing the results for the $\overline{\mbox{DR}}$ scheme is

\begin{equation}\label{Gamma_Constants_DR}
g_{12} = -\frac{1}{2} - \frac{B}{2},
\end{equation}

\noindent
where $B$ is a regularization parameter related to the function $F$ in Eq. (\ref{Action_Regularized_By_HD}),

\begin{equation}\label{B_Definition}
B\equiv \int\limits_0^\infty dx\,\ln x\, \frac{d}{dx}\frac{1}{F^2(x)}.
\end{equation}

\noindent
Comparing Eq. (\ref{Gamma_DR_Bar}) with Eq. (\ref{Gamma_Renormalized_Result}) we see that the result for the  $\overline{\mbox{DR}}$ scheme can be reproduced if the finite constants fixing a renormalization prescription take the values

\begin{equation}\label{Gamma_Constants_DR_Continued}
g_{23} = \frac{1}{2}\Big(C + g_{12}^2 - g_{12} +\frac{1}{4}\Big) = \frac{C}{2} + \frac{(2+B)^2}{8};\qquad g_{24} = \frac{1}{2} g_{12}^2 = \frac{(1+B)^2}{8},
\end{equation}

\noindent
while all scheme-independent terms coincide. Certainly, this confirms correctness of the calculation.

\medskip

\textbf{3)} The \textbf{``minimal'' scheme} is constructed in such a way that RGFs have the simplest possible form, while all relations inherent in supersymmetric theories remain valid. In particular, we admit only such renormalization procedures for which Eq. (\ref{Lambda_Renormalization}) is valid in all orders. This implies that it is impossible to get rid of the term containing $\zeta(3)$ in Eq. (\ref{Gamma_Renormalized_Result}) as it was done in Ref. \cite{Jack:1996qq} (Certainly, the corresponding renormalization procedure is possible, but it breaks Eq. (\ref{Lambda_Renormalization}).)

Looking at Eq. (\ref{Gamma_Renormalized_Result}), we see that it is possible to set all scheme-dependent terms to 0 if the finite constants which determine the renormalization prescription in the two-loop approximation satisfy the constraints

\begin{equation}\label{Gamma_Constants_Minimal}
g_{23} = \frac{1}{2}\Big(C + g_{12}^2 - g_{12}\Big);\qquad g_{24} = \frac{1}{2} g_{12}^2.
\end{equation}

\noindent
In this case the part of the three-loop anomalous dimension under consideration takes the simplest form

\begin{eqnarray}\label{Gamma_Minimal}
&&\hspace*{-5mm} \widetilde\gamma_i{}^j(\alpha,\lambda)\Big|_{\mbox{\scriptsize Minimal}} = \frac{1}{4\pi^2}\lambda^*_{iab} \lambda^{jab} - \frac{1}{16\pi^4} \lambda^*_{iab} \lambda^{jac} \lambda^*_{cde} \lambda^{bde}
+ \frac{3\zeta(3)}{64\pi^6}\,\lambda^{jkl} \lambda^*_{lrs} \lambda^{pmr} \lambda^*_{kmn} \lambda^{qns} \lambda^*_{ipq}
\nonumber\\
&&\hspace*{-5mm} + \frac{1}{32\pi^6}\,\lambda^{jml} (\lambda^*_{mkp} \lambda^{pab} \lambda^*_{qab} \lambda^{nkq}) \lambda^*_{inl} + O(\alpha) + \mbox{higher orders}.\vphantom{\frac{1}{2}}
\end{eqnarray}

\noindent
In particular, we see that the dependence on regularization parameters completely disappeared.

\section{The four-loop contribution to the $\beta$-function}
\hspace*{\parindent}\label{Section_Beta}

According to \cite{Stepanyantz:2020uke}, the NSVZ equation (\ref{Beta_NSVZ}) is valid for RGFs defined in terms of the bare couplings in the case of using the higher covariant derivative regularization independently of a particular renormalization procedure. That is why the four-loop $\beta$-function can immediately be obtained if the anomalous dimension of the matter superfields has been calculated in the three-loop approximation with this regularization. In particular, starting from Eq. (\ref{Gamma_Bare_Result}) we are able to construct a part of the four-loop $\beta$-function containing the leading terms in the Yukawa couplings,

\begin{eqnarray}\label{Beta_Bare}
&&\hspace*{-5mm} \frac{\beta(\alpha_0,\lambda_0)}{\alpha_0^2} = -\frac{1}{2\pi}\Big(3C_2-T(R)\Big) - \frac{1}{2\pi r} C(R)_j{}^i \bigg\{\frac{1}{4\pi^2}\lambda^*_{0iab} \lambda_0^{jab}
- \frac{1}{16\pi^4} \lambda^*_{0iab} \lambda_0^{jac} \lambda^*_{0cde} \lambda_0^{bde}\qquad
\nonumber\\
&&\hspace*{-5mm} + \frac{3\zeta(3)}{64\pi^6}\,\lambda_0^{jkl} \lambda^*_{0lrs} \lambda_0^{pmr} \lambda^*_{0kmn} \lambda_0^{qns} \lambda^*_{0ipq} + \frac{1}{32\pi^6}\,\lambda_0^{jml} (\lambda^*_{0mkp} \lambda_0^{pab} \lambda^*_{0qab} \lambda_0^{nkq}) \lambda^*_{0inl} + \frac{C}{64\pi^6} \nonumber\\
&&\hspace*{-5mm} \times \Big(2\lambda_0^{jmp} (\lambda^*_{0mab} \lambda_0^{nab})(\lambda^*_{0ncd}\lambda_0^{kcd}) \lambda^*_{0ikp}
+ \lambda_0^{jmn} (\lambda^*_{0mab} \lambda_0^{pab}) (\lambda^*_{0ncd} \lambda_0^{qcd}) \lambda^*_{0ipq}\Big)\bigg\} + O(\alpha_0)\vphantom{\frac{1}{2}}\nonumber\\
&&\hspace*{-5mm} + \mbox{higher orders}.\vphantom{\bigg\{}
\end{eqnarray}

Integrating the renormalization group equation for the gauge coupling $\alpha_0$ in the lowest orders we obtain the relation between the bare and renormalized couplings. Maximally following the notations adopted in \cite{Kazantsev:2018nbl,Kazantsev:2020kfl}, this relation can be presented in the form

\begin{eqnarray}\label{Alpha_Relation}
&&\hspace*{-6mm} \frac{1}{\alpha}-\frac{1}{\alpha_0} = -\frac{3}{2\pi} C_2 \Big(\ln\frac{\Lambda}{\mu} + b_{11} \Big) + \frac{1}{2\pi} T(R) \Big(\ln\frac{\Lambda}{\mu} + b_{12} \Big)
-\frac{1}{8\pi^3 r} C(R)_j{}^i \lambda^*_{imn} \lambda^{jmn}\nonumber\\
&&\hspace*{-6mm} \times \Big(\ln\frac{\Lambda}{\mu}+b_{24} \Big)
-\frac{1}{16\pi^5r} C(R)_j{}^i \lambda^*_{iac} \lambda^{jab} \lambda^*_{bde} \lambda^{cde} \Big( \frac{1}{2} \ln^2\frac{\Lambda}{\mu} +g_{12} \ln\frac{\Lambda}{\mu} -\frac{1}{2} \ln\frac{\Lambda}{\mu} +b_{34}\Big)
\nonumber\\
&&\hspace*{-6mm} - \frac{1}{32\pi^5 r} C(R)_j{}^i \lambda^*_{iab} \lambda^{kab} \lambda^*_{kcd} \lambda^{jcd} \Big(\frac{1}{2}
\ln^2\frac{\Lambda}{\mu} +g_{12} \ln\frac{\Lambda}{\mu} +b_{35}\Big) + O(\alpha) + \mbox{higher orders},\vphantom{\frac{1}{2}}
\end{eqnarray}

\noindent
where we included all terms proportional to $\alpha$ into $O(\alpha)$ without writing them out explicitly. Again, we have verified that all coefficients at higher logarithms in the four-loop generalization of this expression in the HD+MSL scheme agree with the general expressions produced by the renormalization group  \cite{Derkachev:2017nhd,Meshcheriakov:2022tyi,Meshcheriakov:2023fmk,Meshcheriakov:2024qwj,Kovyrshin:2025ufp}. (To be precise, for $\ln Z_\alpha = \ln (\alpha/\alpha_0)$ we have verified the validity of Eq. (33) in Ref. \cite{Kovyrshin:2025ufp}.)

Note that for obtaining the four-loop $\beta$-function we in fact need the relation between $\alpha$ and $\alpha_0$ in the next order of the perturbation theory. However, it is rather lengthy and is not presented here because new finite constants appearing in the three-loop order do not enter the four-loop $\beta$-function. As usual, to find the expression for it, we express $1/\alpha$ in terms of $\alpha_0$ and $\lambda_0^{ijk}$ and differentiate the result with respect to $\ln\mu$ at fixed values of the bare couplings. After that, it is necessary to express the result in terms of the renormalized couplings. Then the expression for the part of the four-loop $\beta$-function composed of the leading terms in the Yukawa couplings takes the form

\begin{eqnarray}\label{Beta_Renormalized_Result}
&&\hspace*{-7mm}  \frac{\widetilde\beta(\alpha,\lambda)}{\alpha^2} = - \frac{1}{2\pi} \Big(3C_2-T(R)\Big) - \frac{1}{2\pi r} C(R)_j{}^i\bigg\{ \frac{1}{4\pi^2}\lambda^*_{iab} \lambda^{jab} - \frac{1}{8\pi^4} \lambda^*_{iab} \lambda^{jac} \lambda^*_{cde} \lambda^{bde} \Big(\frac{1}{2}\nonumber\\
&&\hspace*{-7mm} + b_{24} - g_{12}\Big) + \frac{1}{16\pi^4} (\lambda^*_{iab} \lambda^{kab}) (\lambda^*_{kcd} \lambda^{jcd}) \Big(g_{12}-b_{24}\Big) + \frac{3\zeta(3)}{64\pi^6}\,\lambda^{jkl} \lambda^*_{lrs} \lambda^{pmr} \lambda^*_{kmn} \lambda^{qns} \lambda^*_{ipq}
\nonumber\\
&&\hspace*{-7mm}
+ \frac{1}{32\pi^6}\lambda^{jml} (\lambda^*_{mkp} \lambda^{pab} \lambda^*_{qab} \lambda^{nkq}) \lambda^*_{inl}\Big(1-g_{12}+2g_{23}+b_{24}-2b_{34}\Big)
+ \frac{1}{64\pi^6}\lambda^{jmp} (\lambda^*_{mab} \lambda^{nab})\nonumber\\
&&\hspace*{-7mm} \times (\lambda^*_{ncd}\lambda^{kcd}) \lambda^*_{ikp}\Big(2C +g_{12}^{2}+2g_{24} - 2g_{12} - 4b_{34}\Big)
+ \frac{1}{64\pi^6}\lambda^{jmn} (\lambda^*_{mab} \lambda^{pab}) (\lambda^*_{ncd} \lambda^{qcd}) \lambda^*_{ipq}\nonumber\\
&&\hspace*{-7mm}\times \Big(C- g_{12}+g_{12}^{2}-2b_{34}\Big)
+ \frac{1}{64\pi^6}(\lambda^*_{iab}\lambda^{kab})(\lambda^*_{kcd} \lambda^{jce} \lambda^*_{efg} \lambda^{dfg}) \Big(2g_{12}^{2}-g_{12}+2g_{23}+b_{24}\nonumber\\
&&\hspace*{-7mm} -2b_{34}-4b_{35}\Big)
+ \frac{1}{128\pi^6} (\lambda^*_{iab}\lambda^{kab})(\lambda^*_{kcd}\lambda^{lcd})(\lambda^*_{lef}\lambda^{jef})\Big(g_{12}^2 +2g_{24} -4b_{35}\Big)
\bigg\} + O(\alpha) \nonumber\\
&&\hspace*{-7mm} + \mbox{higher orders}. \vphantom{\frac{1}{\pi^2}}
\end{eqnarray}

\noindent
Comparing this expression with Eq. (\ref{Gamma_Renormalized_Result}) we see that the NSVZ equation is valid if the finite constants satisfy the constraints

\begin{equation}\label{Constants_NSVZ}
b_{24} = g_{12};\qquad b_{34} = g_{23};\qquad b_{35} = g_{24}.
\end{equation}

As was previously done for the anomalous dimension, it is expedient to consider three particular cases of this expression corresponding to specific renormalization prescription.

\medskip

\textbf{1)} As usual, the \textbf{HD+MSL scheme} is obtained when all parameters fixing a renormalization prescription are set to 0, in particular,

\begin{equation}\label{Beta_Constants_HD+MSL}
g_{12} = g_{23} = g_{24} = 0;\qquad b_{11} = b_{12} = b_{24} = b_{34} = b_{35} = 0,\qquad \mbox{etc}.
\end{equation}

\noindent
In this case the $\beta$-function defined in terms of the renormalized couplings coincides with the one defined in terms of the bare couplings after the formal replacement $\alpha_0\to\alpha$, $\lambda_0^{ijk}\to \lambda^{ijk}$ and is given by the expression

\begin{eqnarray}\label{Beta_HD+MSL}
&&\hspace*{-5mm} \frac{\widetilde\beta(\alpha,\lambda)}{\alpha^2}\bigg|_{\mbox{\scriptsize HD+MSL}} = -\frac{1}{2\pi}\Big(3C_2-T(R)\Big) - \frac{1}{2\pi r} C(R)_j{}^i \bigg\{\frac{1}{4\pi^2}\lambda^*_{iab} \lambda^{jab}
- \frac{1}{16\pi^4} \lambda^*_{iab} \lambda^{jac} \lambda^*_{cde} \lambda^{bde}
\nonumber\\
&&\hspace*{-5mm} + \frac{3\zeta(3)}{64\pi^6}\,\lambda^{jkl} \lambda^*_{lrs} \lambda^{pmr} \lambda^*_{kmn} \lambda^{qns} \lambda^*_{ipq} + \frac{1}{32\pi^6}\,\lambda^{jml} (\lambda^*_{mkp} \lambda^{pab} \lambda^*_{qab} \lambda^{nkq}) \lambda^*_{inl} + \frac{C}{64\pi^6} \Big(2\lambda^{jmp} (\lambda^*_{mab} \nonumber\\
&&\hspace*{-5mm} \times\, \lambda^{nab})(\lambda^*_{ncd}\lambda^{kcd}) \lambda^*_{ikp}
+ \lambda^{jmn} (\lambda^*_{mab} \lambda^{pab}) (\lambda^*_{ncd} \lambda^{qcd}) \lambda^*_{ipq}\Big)\bigg\} + O(\alpha) + \mbox{higher orders}.\vphantom{\bigg\{}
\end{eqnarray}

\noindent
Although the NSVZ equation is satisfied in this case, RGFs depend on regularization parameters. In particular, the expression (\ref{Beta_HD+MSL}) depends on the regularization-dependent constant $C$. For terms proportional to $\alpha$ the number of similar constants becomes much larger. Certainly, this ambiguity in the choice of regularization parameters is a drawback of the HD+MSL renormalization prescription.

\medskip

\textbf{2)} In the $\overline{\mbox{\textbf{DR}}}$ \textbf{scheme} the four-loop $\beta$-function has been constructed in \cite{Jack:1996cn,Jack:1998uj} with the help of the NSVZ equation and a special finite renormalization.\footnote{For the pure ${\cal N}=1$ supersymmetric Yang--Mills theory this result has been confirmed by the explicit calculation made in \cite{Harlander:2006xq}.} The part of this $\beta$-function containing the leading Yukawa terms (in the notations adopted in this paper) is given by the expression

\begin{eqnarray}\label{Beta_DR_Bar}
&&\hspace*{-7mm}  \frac{\widetilde\beta(\alpha,\lambda)}{\alpha^2}\bigg|_{\overline{\mbox{\scriptsize DR}}}  = - \frac{1}{2\pi} \Big(3C_2-T(R)\Big) - \frac{1}{2\pi r} C(R)_j{}^i\bigg\{ \frac{1}{4\pi^2}\lambda^*_{iab} \lambda^{jab} - \frac{3}{32\pi^4} \lambda^*_{iab} \lambda^{jac} \lambda^*_{cde} \lambda^{bde} \nonumber\\
&&\hspace*{-7mm} - \frac{1}{64\pi^4} (\lambda^*_{iab} \lambda^{kab}) (\lambda^*_{kcd} \lambda^{jcd}) + \frac{3\zeta(3)}{64\pi^6}\,\lambda^{jkl} \lambda^*_{lrs} \lambda^{pmr} \lambda^*_{kmn} \lambda^{qns} \lambda^*_{ipq}
+ \frac{19}{384\pi^6}\lambda^{jml} (\lambda^*_{mkp} \lambda^{pab}
\nonumber\\
&&\hspace*{-7mm}
\times \lambda^*_{qab} \lambda^{nkq}) \lambda^*_{inl}
+ \frac{1}{384\pi^6}\lambda^{jmp} (\lambda^*_{mab} \lambda^{nab}) (\lambda^*_{ncd}\lambda^{kcd}) \lambda^*_{ikp}
+ \frac{1}{768\pi^6}\lambda^{jmn} (\lambda^*_{mab} \lambda^{pab}) (\lambda^*_{ncd} \nonumber\\
&&\hspace*{-7mm}
\times\lambda^{qcd}) \lambda^*_{ipq} + \frac{5}{768\pi^6}(\lambda^*_{iab}\lambda^{kab})(\lambda^*_{kcd} \lambda^{jce} \lambda^*_{efg} \lambda^{dfg})
- \frac{1}{768\pi^6} (\lambda^*_{iab}\lambda^{kab})(\lambda^*_{kcd}\lambda^{lcd})(\lambda^*_{lef}\lambda^{jef})
\bigg\} \nonumber\\
&&\hspace*{-7mm} + O(\alpha) + \mbox{higher orders}. \vphantom{\frac{1}{\pi^2}}
\end{eqnarray}

\noindent
Comparing this expression with Eq. (\ref{Beta_Renormalized_Result}) and taking into account Eqs. (\ref{Gamma_Constants_DR}) and (\ref{Gamma_Constants_DR_Continued}), we obtain the values of finite constants for which Eq. (\ref{Beta_Renormalized_Result}) reproduces the $\overline{\mbox{DR}}$ result,\footnote{Values of the coefficients $b_{11}$ and $b_{12}$ corresponding to the $\overline{\mbox{DR}}$ scheme were found in \cite{Kazantsev:2020kfl}. They are related to the masses of the Pauli--Villars superfields which are used for regularizing residual one-loop divergences and are given by the expressions $b_{11}=\ln M_\varphi/\Lambda$ and $b_{12}=\ln M/\Lambda$.}

\begin{eqnarray}\label{Beta_Constants_DR}
&& b_{24} = g_{12} + \frac{1}{4} = -\frac{1}{4} - \frac{B}{2};\qquad \nonumber\\
&& b_{34} = g_{23} - \frac{1}{6} = \frac{1}{2} g_{12}^2 - \frac{1}{2}g_{12} + \frac{C}{2} - \frac{1}{24} = \frac{C}{2} +\frac{(2+B)^2}{8} -\frac{1}{6};\quad\nonumber\\
&& b_{35} = g_{24} + \frac{1}{24} = \frac{1}{2} g_{12}^2 + \frac{1}{24} = \frac{(1+B)^2}{8} + \frac{1}{24}.
\end{eqnarray}

\noindent
Note that the very existence of such values that give Eq. (\ref{Beta_DR_Bar}) is rather nontrivial and confirms the correctness of the calculation and of the results derived in \cite{Jack:1996cn}.

Although the expression (\ref{Beta_DR_Bar}) is simple and unambiguous, it does not satisfy the NSVZ relation. (This can be easily seen by comparing Eqs. (\ref{Constants_NSVZ}) and (\ref{Beta_Constants_DR}).) Therefore, for supersymmetric theories it seems better to use a different renormalization prescription for which all equations inherent in supersymmetric theories are valid.

\medskip

\textbf{3)} By definition, in the \textbf{``minimal'' scheme} the NSVZ relation should be satisfied. Therefore, the four-loop $\beta$-function can be immediately obtained by substituting the three-loop anomalous dimension (\ref{Gamma_Minimal}) into the NSVZ equation. For the part of it discussed in this paper, the result takes the form

\begin{eqnarray}\label{Beta_Minimal}
&&\hspace*{-5mm} \frac{\widetilde\beta(\alpha,\lambda)}{\alpha^2}\bigg|_{\mbox{\scriptsize Minimal}} =  -\frac{1}{2\pi}\Big(3C_2-T(R)\Big) - \frac{1}{2\pi r} C(R)_j{}^i \bigg\{\frac{1}{4\pi^2}\lambda^*_{iab} \lambda^{jab}
- \frac{1}{16\pi^4} \lambda^*_{iab} \lambda^{jac} \lambda^*_{cde}
\nonumber\\
&&\hspace*{-5mm}\times \lambda^{bde} + \frac{3\zeta(3)}{64\pi^6} \lambda^{jkl} \lambda^*_{lrs} \lambda^{pmr} \lambda^*_{kmn} \lambda^{qns} \lambda^*_{ipq} + \frac{1}{32\pi^6}\,\lambda^{jml} (\lambda^*_{mkp} \lambda^{pab} \lambda^*_{qab} \lambda^{nkq}) \lambda^*_{inl} \bigg\} + O(\alpha)\nonumber\\
&&\hspace*{-5mm} + \mbox{higher orders}.\vphantom{\frac{1}{2}}
\end{eqnarray}

\noindent
Comparing this expression with Eq. (\ref{Beta_Renormalized_Result}), we see that the corresponding values of the finite constants (some of them are taken from Eq. (\ref{Gamma_Constants_Minimal})) are given by

\begin{equation}\label{Beta_Constants_Minimal}
b_{24} = g_{12};\qquad b_{34} = g_{23} = \frac{1}{2}\big(C + g_{12}^2 - g_{12}\big);\qquad b_{35} = g_{24}= \frac{1}{2} g_{12}^2.
\end{equation}

\noindent
This in particular implies that, at least, in the approximation under consideration and for the terms of the structure in question, the minimal scheme exists and is unique.

\section{Conclusion}
\hspace*{\parindent}

In this paper we investigated certain scheme-dependent quantum corrections in ${\cal N}=1$ supersymmetric gauge theories. Namely, we calculated the three-loop contribution to the anomalous dimension of the matter superfields proportional to the sixth power of the Yukawa couplings in the case of using the Slavnov's higher covariant derivative regularization supplemented by an arbitrary renormalization prescription compatible with Eq. (\ref{Lambda_Renormalization}). This choice of the regularization is motivated by the fact that in this case the all-loop NSVZ scheme can be naturally constructed and is given by the HD+MSL prescription \cite{Kataev:2013eta,Stepanyantz:2020uke}, while in the $\overline{\mbox{DR}}$ scheme the NSVZ equation does not hold \cite{Jack:1996vg,Jack:1996cn,Jack:1998uj}. Therefore, knowing the anomalous dimension in a certain loop, one can easily construct a $\beta$-function in the next loop without calculating the corresponding Feynman (super)diagrams.

The three-loop contribution to the anomalous dimension proportional to the sixth power of Yukawa couplings is given by the expression (\ref{Gamma_Renormalized_Result}). This expression contains some terms which depend on the regularization parameters and the parameters fixing a renormalization prescription. The presence of these parameters significantly complicates the result and makes it difficult to analyze. Moreover, the loop integrals which determine scheme-dependent terms are particularly difficult to calculate in the case of using the higher covariant derivative regularization. That is why in this paper we focused on the question of how to construct such a renormalization prescription in which RGFs have the simplest form, but all supersymmetric equations relating the gauge and Yukawa $\beta$-functions to the anomalous dimension of the matter superfields remain valid. We call this scheme ``minimal''. The results for the parts of the three-loop anomalous dimension and the four-loop $\beta$-function containing terms with the largest powers of Yukawa couplings in the minimal scheme are given by Eqs. (\ref{Gamma_Minimal}) and (\ref{Beta_Minimal}), respectively. They are  very simple (even simpler than the corresponding $\overline{\mbox{DR}}$ results given by Eqs. (\ref{Gamma_DR_Bar}) and (\ref{Beta_DR_Bar})), satisfy the NSVZ equation, and do not contain arbitrary parameters. Therefore, RGFs in the minimal scheme appear to be unambiguously defined and regularization independent, at least, for the terms in question. It should be noted that the multiloop integrals which determine these terms can be calculated even in the case of using the higher covariant derivative regularization. That is why the use of the minimal scheme allows one to get rid of the calculations of the most complicated loop integrals. Certainly, it can considerably simplify the multiloop calculations and help to extract the simplest result that keeps all features inherent in supersymmetric theories.

However, some open questions still remain. In particular, it is necessary to reveal whether the minimal scheme really exists and is unambiguous in all orders of the perturbation theory in the non-Abelian case. Moreover, it would be interesting to investigate the form of RGFs in this scheme in all orders for certain theories (e.g., for MSSM) and their general features. The minimal scheme may also be useful for understanding some interesting properties of special classes of supersymmetric theories, like the $P=\frac{1}{3}Q$ theories \cite{Jack:1995gm,Jack:1996qq}. We hope to investigate these issues in future studies.

\appendix

\section*{Appendix}

\section{Calculation of the two-loop integral}
\label{Appendix_Two-Loop}
\hspace*{\parindent}

Let us first describe the calculation of the two-loop integral present in Eq. (\ref{Gamma_2Loop_Integrals}). To simplify the expression for it, we first consider the integral

\begin{equation}
\int \frac{d^4K}{(2\pi)^4}  \frac{d^4L}{(2\pi)^4} \frac{8}{K^4 F_K^2 L^2 F_L (K+L)^2} \Big(\frac{1}{F_K F_{K+L}} -\frac{1}{F_L}\Big),
\end{equation}

\noindent
which is dimensionless and convergent both in the ultraviolet and infrared regions. Therefore, it is a finite constant independent of $\Lambda$. That is why its derivative with respect to $\Lambda$ vanishes. This implies that the two-loop expression for the (considered part of the) anomalous dimension can be rewritten in the form

\begin{eqnarray}\label{Gamma_2Loop_Integrals_Modified}
&&\hspace*{-5mm} \gamma_i{}^j(\alpha_0,\lambda_0) = \frac{1}{4\pi^2}\lambda^*_{0iab} \lambda_0^{jab}
- \lambda^*_{0iab} \lambda_0^{jac} \lambda^*_{0cde} \lambda_0^{bde} \int \frac{d^4K}{(2\pi)^4}  \frac{d^4L}{(2\pi)^4} \Big(\frac{8}{K^4 L^2 (K+L)^2} - \frac{4}{K^4 L^4}\Big)\nonumber\\
&&\hspace*{-5mm} \times \frac{d}{d\ln\Lambda}\Big(\frac{1}{F_K^2 F_L^2}\Big) + O(\alpha_0) +\mbox{higher orders}.
\end{eqnarray}

\noindent
For calculating the integral in this equation, we will use the method \cite{Rosner:1967zz} based on the Chebyshev polynomials

\begin{equation}
C_n(\cos\theta) \equiv \frac{\sin\, ((n+1)\theta)}{\sin\theta}.
\end{equation}

\noindent
These polynomials satisfy the equation

\begin{equation}
\frac{1}{1-2tz + t^2} = \sum\limits_{n=0}^\infty t^n C_n(z)
\end{equation}

\noindent
valid for $|t|<1$. Due to this equation, the function $(K-L)^{-2} = (K^2 - 2 KL\cos\theta+L^2)^{-1}$ can be written as

\begin{equation}\label{Expansion_Of_(K-L)^2}
\frac{1}{(K-L)^2} = \left\{\begin{array}{l}
{\displaystyle \frac{1}{K^2} \sum\limits_{n=0}^\infty \Big(\frac{L}{K}\Big)^n C_n(\cos\theta),\quad\mbox{for}\quad K > L;}\\
\vphantom{1}\\
{\displaystyle \frac{1}{L^2} \sum\limits_{n=0}^\infty \Big(\frac{K}{L}\Big)^n C_n(\cos\theta),\quad\, \mbox{for}\quad L > K,}
\end{array}
\right.
\end{equation}

\noindent
where $K$ and $L$ are the magnitudes of the Euclidean four-vectors $K_\mu$ and $L_\mu$, respectively, and $\theta$ is the angle between them. Moreover, the identities

\begin{eqnarray}\label{Identity_Product}
&& \int \frac{d\Omega_Q}{2\pi^2} C_m\Big(\frac{K_\mu Q_\mu}{KQ}\Big) C_n\Big(\frac{Q_\nu L_\nu}{QL}\Big) = \frac{1}{n+1} \delta_{mn} C_n\Big(\frac{K_\mu L_\mu}{KL}\Big);\\
\label{Identity_Ortogonality}
&& \int \frac{d\Omega}{2\pi^2} C_m(\cos \theta) C_n(\cos \theta) = \delta_{mn},
\end{eqnarray}

\noindent
where a solid angle on a sphere $S^3$ in the momentum space with the Cartesian coordinates $Q_\mu$ is denoted by $d\Omega_Q$, allow calculating the angular part of loop integrals. Using the identities (\ref{Expansion_Of_(K-L)^2}), (\ref{Identity_Product}), and (\ref{Identity_Ortogonality}) we present the integral which determines the two-loop anomalous dimension (\ref{Gamma_2Loop_Integrals_Modified}) in the form

\begin{eqnarray}
&& I_0 \equiv \int \frac{d^4K}{(2\pi)^4}  \frac{d^4L}{(2\pi)^4} \Big(\frac{8}{K^4 L^2 (K+L)^2} - \frac{4}{K^4 L^4}\Big) \frac{d}{d\ln\Lambda}\Big(\frac{1}{F_K^2 F_L^2}\Big) \qquad\nonumber\\
&&\qquad\qquad\qquad\qquad\qquad\qquad\qquad\qquad\quad = \frac{1}{8\pi^4} \int\limits_0^\infty \frac{dK}{K^3} \int\limits_0^K dL\,L\,\frac{d}{d\ln\Lambda}\Big(\frac{1}{F_K^2 F_L^2}\Big).\qquad
\end{eqnarray}

\noindent
Next, we make the change of variable $L=xK$ and take into account that

\begin{equation}
\frac{d}{d\ln\Lambda}\Big(\frac{1}{F_K^2 F_{xK}^2}\Big) = - K\,\frac{d}{dK} \Big(\frac{1}{F_K^2 F_{xK}^2}\Big).
\end{equation}

\noindent
After that, using the boundary conditions $F(0)=1$, $F(\infty) = \infty$, it is possible to calculate the integral in question for an arbitrary form of the regulator function $F$,

\begin{eqnarray}
&& I_0 = - \frac{1}{8\pi^4} \int\limits_0^\infty dK \int\limits_0^1 dx\,x\,\frac{d}{dK} \Big(\frac{1}{F_K^2 F_{xK}^2}\Big) = \frac{1}{8\pi^4} \int\limits_0^1 dx\,x = \frac{1}{16\pi^4}.
\end{eqnarray}

\noindent
Substituting this expression into Eq. (\ref{Gamma_2Loop_Integrals_Modified}), we obtain the result (\ref{Gamma_2Loop_Result}).

\section{Explicit expressions for the diagrams presented in Fig. \ref{Figure_Lambda6}}
\label{Appendix_Diagrams}
\hspace*{\parindent}

In this appendix we present the explicit expressions for contributions of all supergraphs presented in Fig. \ref{Figure_Lambda6} to the function $G$ defined by Eq. (\ref{G_Function_Definition}). The results are written in the limit of the vanishing external momentum in the Euclidean space after the Wick rotation. Certainly, in this limit each particular expression is not well-defined. However, they can be used for constructing the well-defined expression for the anomalous dimension given by Eq. (\ref{Lambda_Equation}). Denoting the contribution of a diagram with the number $(k)$ by $(\Delta G^{(k)})_i{}^j$, the results can be written as

\begin{eqnarray}
&&\hspace*{-5mm} (\Delta G^{(1)})_i{}^j\Big|_{p\to 0} = 16 \lambda_0^{jkl} \lambda^*_{0lrs} \lambda_0^{pmr} \lambda^*_{0kmn} \lambda_0^{qns} \lambda^*_{0ipq}\, \int \frac{d^4Q}{(2\pi)^4}\,\frac{d^4K}{(2\pi)^4}\, \frac{d^4L}{(2\pi)^4}\,
\frac{1}{K^2 F_K^2 L^2 F_L Q^2 F_Q}\nonumber\\
&&\hspace*{-5mm} \times \frac{1}{(Q+K+L)^2 F_{Q+K+L}^2 (K+L)^2 F_{K+L} (K+Q)^2 F_{K+Q}};\qquad\\
&& \hspace*{-5mm} (\Delta G^{(2)})_i{}^j\Big|_{p\to 0} =  8 \lambda_0^{jmn} (\lambda^*_{0mab} \lambda_0^{pab}) (\lambda^*_{0ncd} \lambda_0^{qcd}) \lambda^*_{0ipq} \int \frac{d^4Q}{(2\pi)^4}\,\frac{d^4K}{(2\pi)^4}\, \frac{d^4L}{(2\pi)^4}\, \frac{1}{K^4 F_K^4 L^2 F_L Q^2 F_Q}\nonumber\\
&&\hspace*{-5mm} \times \frac{1}{(K+L)^2 F_{K+L} (K+Q)^2 F_{k+Q}};\\
&& \hspace*{-5mm} (\Delta G^{(3)})_i{}^j\Big|_{p\to 0} = 16 \lambda_0^{jmp} (\lambda^*_{0mab} \lambda_0^{nab})(\lambda^*_{0ncd}\lambda_0^{kcd}) \lambda^*_{0ikp} \int \frac{d^4Q}{(2\pi)^4}\,\frac{d^4K}{(2\pi)^4}\, \frac{d^4L}{(2\pi)^4}\,
\frac{1}{K^4 F_K^4 L^2 F_L Q^2 F_Q}\nonumber\\
&&\hspace*{-5mm} \times \frac{1}{(K+L)^2 F_{K+L}(K+Q)^2 F_{K+Q}};\\
&& \hspace*{-5mm} (\Delta G^{(4)})_i{}^j\Big|_{p\to 0} = 32 \lambda_0^{jml} (\lambda^*_{0mkp} \lambda_0^{pab} \lambda^*_{0qab} \lambda_0^{nkq}) \lambda^*_{0inl} \int \frac{d^4Q}{(2\pi)^4}\,\frac{d^4K}{(2\pi)^4}\, \frac{d^4L}{(2\pi)^4}\, \frac{1}{K^4 F_K^3 L^2 F_L^2 Q^2 F_Q}\nonumber\\
&&\hspace*{-5mm}\times \frac{1}{(K+L)^2 F_{K+L} (Q+L)^2 F_{Q+L}}.
\end{eqnarray}

\section{Calculation of the three-loop integrals}
\label{Appendix_Three-Loop}
\hspace*{\parindent}

In this appendix we describe the calculation of integrals present in Eq. (\ref{Gamma_3Loop}). The integral $I_1$ (given by Eq. (\ref{Integral_I1})) has already been calculated in \cite{Kuzmichev:2023zxy} for an arbitrary regulator function $F$ and is not considered here. The next integral

\begin{eqnarray}
&&\hspace*{-7mm} I_2 \equiv 32 \int \frac{d^4Q}{(2\pi)^4}\, \frac{d^4K}{(2\pi)^4}\, \frac{d^4L}{(2\pi)^4}\, \frac{d}{d\ln\Lambda}\bigg\{
\frac{1}{K^4 F_K^3 L^2 F_L^2 (K+L)^2 F_{K+L} Q^2 F_Q (Q+L)^2 F_{Q+L}}\nonumber\\
&&\hspace*{-7mm} - \frac{1}{Q^4 F_Q^2 K^4 F_K^3 L^2 F_L (K+L)^2 F_{K+L}} + \frac{1}{3 Q^4 F_Q^2 K^4 F_K^2 L^4 F_L^2} \bigg\}
\end{eqnarray}

\noindent
can also be calculated for an arbitrary regulator function with the help of the technique based on the Chebyshev polynomials. To do this, we should first get rid of $F_{K+L}$ and $F_{Q+L}$ in the denominators. For this purpose, we present the integral under consideration in the form

\begin{equation}
I_2 \equiv I_2'+\Delta I_2,
\end{equation}

\noindent
where

\begin{eqnarray}\label{I2_Prime}
&&\hspace*{-9mm} I_2' \equiv 32 \int \frac{d^4Q}{(2\pi)^4}\, \frac{d^4K}{(2\pi)^4}\, \frac{d^4L}{(2\pi)^4}\, \frac{d}{d\ln\Lambda}\bigg\{
\frac{1}{K^4 F_K^3 L^2 F_L^3 (K+L)^2 Q^2 F_Q^2 (Q+L)^2}\nonumber\\
&&\hspace*{-9mm}\qquad\qquad\qquad\qquad\qquad\qquad\quad - \frac{1}{Q^4 F_Q^2 K^4 F_K^3 L^2 F_L^2 (K+L)^2} + \frac{1}{3 Q^4 F_Q^2 K^4 F_K^2 L^4 F_L^2} \bigg\};\\
\label{Delta_I2}
&&\hspace*{-9mm} \Delta I_2 = 32 \int \frac{d^4Q}{(2\pi)^4}\, \frac{d^4K}{(2\pi)^4}\, \frac{d^4L}{(2\pi)^4}\, \frac{d}{d\ln\Lambda}\bigg\{
\frac{1}{K^4 F_K^3 L^2 F_L^2 (K+L)^2 Q^2 F_Q (Q+L)^2}\nonumber\\
&&\hspace*{-9mm}\qquad\qquad\ \times\Big(\frac{1}{F_{K+L} F_{Q+L}}-\frac{1}{F_L F_Q}\Big) - \frac{1}{Q^4 F_Q^2 K^4 F_K^3 L^2 F_L (K+L)^2}\Big(\frac{1}{F_{K+L}} - \frac{1}{F_L}\Big) \bigg\}.
\end{eqnarray}

\noindent
Let us demonstrate that $\Delta I_2=0$ and, therefore, $I_2=I_2'$. Really, to calculate the first term in Eq. (\ref{Delta_I2}), we note that for any nonsingular function $f(K/\Lambda)$ rapidly decreasing at infinity

\begin{equation}\label{Auxiliary_Identity}
\int \frac{d^4K}{(2\pi)^4} \frac{1}{K^4} \frac{df(K/\Lambda)}{d\ln\Lambda} = -\frac{1}{8\pi^2}\int\limits_0^\infty dK \frac{df(K/\Lambda)}{dK} = \frac{1}{8\pi^2} f(0).
\end{equation}

\noindent
The integral over $d^4Q$ in the second term can also be calculated with the help of a similar equation if we note that

\begin{equation}
\int \frac{d^4K}{(2\pi)^4}\, \frac{d^4L}{(2\pi)^4}\,\frac{1}{K^4 F_K^3 L^2 F_L (K+L)^2}\Big(\frac{1}{F_{K+L}} - \frac{1}{F_L}\Big)
\end{equation}

\noindent
is a finite constant (and, therefore, does not depend on $\Lambda$). Therefore, the expression for $\Delta I_2$ takes the form

\begin{eqnarray}
&& \Delta I_2 = \frac{4}{\pi^2} \int \frac{d^4Q}{(2\pi)^4}\, \frac{d^4L}{(2\pi)^4}\, \frac{1}{L^4 F_L^3 Q^2 F_Q (Q+L)^2} \Big(\frac{1}{F_{Q+L}}-\frac{1}{F_Q}\Big)
\nonumber\\
&&\qquad\qquad\qquad - \frac{4}{\pi^2} \int \frac{d^4K}{(2\pi)^4}\, \frac{d^4L}{(2\pi)^4}\,  \frac{1}{K^4 F_K^3 L^2 F_L (K+L)^2}\Big(\frac{1}{F_{K+L}} - \frac{1}{F_L}\Big) = 0.\qquad
\end{eqnarray}

The remaining integral $I_2'$ can be calculated with the help of the Chebyshev polynomials. Using Eqs. (\ref{Expansion_Of_(K-L)^2}), (\ref{Identity_Product}), and (\ref{Identity_Ortogonality}) we present it in the form

\begin{eqnarray}
&&\hspace*{-9mm} I_2 = I_2' = \frac{32}{(8\pi^2)^3} \Bigg\{
\int\limits_0^\infty \frac{dK}{K^3} \int\limits_0^K \frac{dL}{L} \int\limits_0^L dQ\, Q\, \frac{d}{d\ln\Lambda}\Big(\frac{1}{F_K^3 F_L^3 F_Q^2}\Big)
\nonumber\\
&&\hspace*{-9mm} + \int\limits_0^\infty \frac{dK}{K^3} \int\limits_0^K dL\,L \int\limits_L^\infty \frac{dQ}{Q}\, \frac{d}{d\ln\Lambda}\Big(\frac{1}{F_K^3 F_L^3 F_Q^2}\Big)
+ \int\limits_0^\infty \frac{dK}{K} \int\limits_K^\infty \frac{dL}{L^3} \int\limits_0^L dQ\,Q\, \frac{d}{d\ln\Lambda}\Big(\frac{1}{F_K^3 F_L^3 F_Q^2}\Big)
\nonumber\\
&&\hspace*{-9mm} + \int\limits_0^\infty \frac{dK}{K} \int\limits_K^\infty \frac{dL}{L} \int\limits_L^\infty \frac{dQ}{Q}\, \frac{d}{d\ln\Lambda}\Big(\frac{1}{F_K^3 F_L^3 F_Q^2}\Big)
- \int\limits_0^\infty \frac{dQ}{Q} \int\limits_0^\infty \frac{dK}{K^3} \int\limits_0^K dL\,L\, \frac{d}{d\ln\Lambda}\Big(\frac{1}{F_K^3 F_L^2 F_Q^2}\Big)
\nonumber\\\
&&\hspace*{-9mm} - \int\limits_0^\infty \frac{dQ}{Q} \int\limits_0^\infty \frac{dK}{K} \int\limits_K^\infty \frac{dL}{L}\, \frac{d}{d\ln\Lambda}\Big(\frac{1}{F_K^3 F_L^2 F_Q^2}\Big)
+ \frac{1}{3} \int\limits_0^\infty \frac{dQ}{Q} \int\limits_0^\infty \frac{dK}{K} \int\limits_0^\infty \frac{dL}{L}\, \frac{d}{d\ln\Lambda}\Big(\frac{1}{F_K^2 F_L^2 F_Q^2}\Big) \Bigg\}.
\end{eqnarray}

\noindent
Dividing the integration domain properly and making relevant changes of variables, this expression can be rewritten as

\begin{eqnarray}\label{I2_Intermediate}
&& I_2 = \frac{1}{16\pi^6} \Bigg\{\int\limits_0^\infty \frac{dQ}{Q} \int\limits_0^Q \frac{dK}{K} \int\limits_0^K \frac{dL}{L}\,\frac{d}{d\ln\Lambda}\bigg[\frac{(1-F_L)}{F_Q^2 F_L^3}
\bigg(\frac{1}{F_K^3} - \frac{2}{F_K^2} \bigg)\bigg]\nonumber\\
&& + \int\limits_0^\infty \frac{dQ}{Q^3} \int\limits_0^Q dK\, K \int\limits_0^K \frac{dL}{L}\, \frac{d}{d\ln\Lambda} \bigg[\frac{1-F_L}{F_Q^3 F_K^2 F_L^3}\bigg]
+ \int\limits_0^\infty \frac{dQ}{Q} \int\limits_0^Q \frac{dK}{K^3} \int\limits_0^K dL\,L\, \frac{d}{d\ln\Lambda} \bigg[\frac{1-F_L}{F_Q^2 F_K^3 F_L^3}\bigg]\qquad\nonumber\\
&& + \int\limits_0^\infty \frac{dQ}{Q^3} \int\limits_0^Q \frac{dK}{K} \int\limits_0^K dL\,L\,\frac{d}{d\ln\Lambda}\bigg[\frac{2}{F_Q^3 F_K^3 F_L^2} + \frac{1-F_L}{F_Q^3 F_K^2 F_L^3}\bigg]\Bigg\}.
\end{eqnarray}

\noindent
Next, we make the change of variables $L=yK$ and, after that, $K=xQ$. Moreover, we take into account that for any dimensionless function $f$ depending on $Q/\Lambda$, $x$, and $y$

\begin{equation}
\frac{d}{d\ln\Lambda} f(Q/\Lambda,x,y) = - Q \frac{d}{dQ} f(Q/\Lambda,x,y).
\end{equation}

\noindent
Then the integral (\ref{I2_Intermediate}) takes the form

\begin{eqnarray}
&& I_2 = - \frac{1}{16\pi^6} \int\limits_0^\infty dQ \int\limits_0^1 dx \int\limits_0^1 dy\, \frac{d}{dQ}\bigg[\frac{(1-F_{xyQ})}{xy F_Q^2 F_{xyQ}^3}
\bigg(\frac{1}{F_{xQ}^3} - \frac{2}{F_{xQ}^2} \bigg) + \frac{x(1-F_{xyQ})}{y F_Q^3 F_{xQ}^2 F_{xyQ}^3}\nonumber\\
&& + \frac{y(1-F_{xyQ})}{x F_Q^2 F_{xQ}^3 F_{xyQ}^3}+\frac{2xy}{F_Q^3 F_{xQ}^3 F_{xyQ}^2} + \frac{xy(1-F_{xyQ})}{F_Q^3 F_{xQ}^2 F_{xyQ}^3}\bigg].
\end{eqnarray}

\noindent
Evidently, the integral over $dQ$ can be easily calculated. Taking into account that $F(0)=1$ and $F$ rapidly increases at large values of its argument, we see that the resulting expression appears to be independent of a particular form of this function,

\begin{equation}
I_2 = \frac{1}{8\pi^6} \int\limits_0^1 dx\, x \int\limits_0^1 dy\,y = \frac{1}{32\pi^6}.
\end{equation}

\medskip

Unfortunately, the remaining integral in Eq. (\ref{Gamma_3Loop}),

\begin{eqnarray}\label{I3_Appendix}
&&\hspace*{-5mm} I_3 \equiv \int\frac{d^4Q}{(2\pi)^4}\, \frac{d^4K}{(2\pi)^4}\, \frac{d^4L}{(2\pi)^4}\, \bigg\{ \frac{d}{d\ln\Lambda}\frac{8}{K^4F_K^4 L^2 F_L Q^2 F_Q (K+L)^2 F_{K+L} (K+Q)^2 F_{K+Q}}\nonumber\\
&&\hspace*{-5mm} - \frac{16}{K^4 F_K^3 L^2 F_L (K+L)^2 F_{K+L}}\cdot \frac{d}{d\ln\Lambda} \frac{1}{Q^4 F_Q^2}\bigg\},
\end{eqnarray}

\noindent
cannot be calculated for an arbitrary higher derivative regulator function $F$. However, it can be simplified to such an extent that it will be obvious that it is a finite constant. For this purpose, first, it is convenient to introduce the auxiliary integral

\begin{equation}\label{Auxiliary_Integral}
I(K/\Lambda)\equiv \int\frac{d^4L}{(2\pi)^4} \frac{1}{L^2 F_L (K+L)^2 F_{K+L}} = \frac{1}{8\pi^2}\bigg[ \ln\frac{\Lambda}{K} + \frac{1}{2} (1-B) + f(K^2/\Lambda^2) \bigg],
\end{equation}

\noindent
where the finite constant $B$ is given by Eq. (\ref{B_Definition}) and $f(K^2/\Lambda^2) = O(K^2/\Lambda^2)$ as $K$ approaches 0. (The last equality in Eq. (\ref{Auxiliary_Integral}) was proven in \cite{Kazantsev:2020kfl}.) Next, we note that the dimensionless integral

\begin{equation}
\int \frac{d^4K}{(2\pi)^4} \frac{(1-F_K)}{K^4 F_K^4} I(K/\Lambda)^2
\end{equation}

\noindent
is convergent in both ultraviolet and infrared regions. Therefore, it is a finite constant, and its derivative with respect to $\ln\Lambda$ vanishes. Taking this into account and involving Eq. (\ref{One-Loop_Integral}) for calculating the integral over $d^4Q$ in the second term of Eq. (\ref{I3_Appendix}), we rewrite the integral $I_3$ in the form

\begin{eqnarray}
&& I_3 = 8\int\frac{d^4K}{(2\pi)^4} \frac{1}{K^4} \bigg[ I(K/\Lambda)^2 \frac{d}{d\ln\Lambda}\frac{1}{F_K^3} + \frac{2}{F_K^3} I(K/\Lambda) \Big(\frac{dI(K/\Lambda)}{d\ln\Lambda} - \frac{1}{8\pi^2}\Big)
\bigg]\nonumber\\
&&\qquad\qquad\qquad\qquad\qquad\quad = 8 \int \frac{d^4K}{(2\pi)^4} \frac{1}{K^4} \bigg[I(K/\Lambda)^2\frac{d}{d\ln\Lambda}\frac{1}{F_K^3} +\frac{I(K/\Lambda)}{4\pi^2 F_K^3}\frac{df(K^2/\Lambda^2)}{d\ln\Lambda}\bigg].\qquad
\end{eqnarray}

\noindent
Substituting the expression (\ref{Auxiliary_Integral}) into $I_3$ and converting the derivative with respect to $\ln\Lambda$ into the derivative with respect to $\ln K$ with the opposite sign, we obtain

\begin{eqnarray}
&& I_3 = - \frac{1}{8\pi^4} \int \frac{d^4K}{(2\pi)^4} \frac{1}{K^4} \bigg[\Big(\ln\frac{\Lambda}{K} + \frac{1-B}{2}\Big)^2\frac{d}{d\ln K} \Big(\frac{1}{F_K^3}\Big)
\nonumber\\
&&\qquad\qquad\qquad\qquad + 2\Big(\ln\frac{\Lambda}{K} + \frac{1-B}{2}\Big) \frac{d}{d\ln K}\Big(\frac{f(K^2/\Lambda^2)}{F_K^3}\Big) + \frac{d}{d\ln K}\Big(\frac{f^2(K^2/\Lambda^2)}{F_K^3}\Big)\bigg].\qquad
\end{eqnarray}

\noindent
It is easy to see that the last term in this expression vanishes, while the others, after the integration by parts in the second term and the change of variable $x=K^2/\Lambda^2$, give the result

\begin{equation}
I_3 = - \frac{1}{256\pi^6} \int\limits_0^\infty dx \Big(\ln x+ B-1\Big)^2 \frac{d}{dx}\Big(\frac{1}{F^3(x)}\Big) - \frac{1}{64\pi^6}\int\limits_0^\infty \frac{dx\,f(x)}{x F^3(x)}.
\end{equation}

\noindent
Evidently, both integrals in this expression converge as $x$ approaches both 0 and infinity. The convergence at infinity is ensured by the rapid growth of the regulator function $F$, while the convergence at 0 follows from the fact that, as $x$ approaches 0, the functions $f(x)$ and $F'(x)$ behave like positive powers of $x$. It is also evident that the result for $I_3$ depends on a particular choice of the function $F$. Therefore, the integral $I_3$ is a regularization dependent finite constant, unlike the integral $I_2$, which does not depend on regularization.


\begin{thebibliography}{150}

%\cite{Grisaru:1979wc}
\bibitem{Grisaru:1979wc}
M.~T.~Grisaru, W.~Siegel and M.~Rocek,
%``Improved Methods for Supergraphs,''
Nucl. Phys. B \textbf{159} (1979), 429.
%doi:10.1016/0550-3213(79)90344-4

%\cite{Novikov:1983uc}
\bibitem{Novikov:1983uc}
V.~A.~Novikov, M.~A.~Shifman, A.~I.~Vainshtein and V.~I.~Zakharov,
%``Exact Gell-Mann-Low Function of Supersymmetric Yang-Mills Theories from Instanton Calculus,''
Nucl. Phys. B \textbf{229} (1983), 381.
%doi:10.1016/0550-3213(83)90338-3

%\cite{Jones:1983ip}
\bibitem{Jones:1983ip}
D.~R.~T.~Jones,
%``More on the Axial Anomaly in Supersymmetric {Yang-Mills} Theory,''
Phys. Lett.  \textbf{123B} (1983), 45.
%doi:10.1016/0370-2693(83)90955-3

%\cite{Novikov:1985rd}
\bibitem{Novikov:1985rd}
V.~A.~Novikov, M.~A.~Shifman, A.~I.~Vainshtein and V.~I.~Zakharov,
%``Beta Function in Supersymmetric Gauge Theories: Instantons Versus Traditional Approach,''
Phys. Lett.  \textbf{166B} (1986), 329
[Sov. J. Nucl. Phys.  \textbf{43} (1986), 294]
[Yad. Fiz. \textbf{43} (1986), 459].
%doi:10.1016/0370-2693(86)90810-5

%\cite{Shifman:1986zi}
\bibitem{Shifman:1986zi}
M.~A.~Shifman and A.~I.~Vainshtein,
%``Solution of the Anomaly Puzzle in SUSY Gauge Theories and the Wilson Operator Expansion,''
Nucl. Phys. B \textbf{277} (1986), 456
[Sov. Phys. JETP \textbf{64} (1986), 428]
[Zh. Eksp. Teor. Fiz.  \textbf{91} (1986), 723].
%doi:10.1016/0550-3213(86)90451-7

%\cite{Parkes:1984dh}
\bibitem{Parkes:1984dh}
A.~Parkes and P.~C.~West,
%``Finiteness in Rigid Supersymmetric Theories,''
Phys. Lett. B \textbf{138} (1984), 99.%-104
%doi:10.1016/0370-2693(84)91881-1

%\cite{Kazakov:1986bs}
\bibitem{Kazakov:1986bs}
D.~I.~Kazakov,
%``Finite $N=1$ {SUSY} Field Theories and Dimensional Regularization,''
Phys. Lett. B \textbf{179} (1986), 352.%-354
%doi:10.1016/0370-2693(86)90491-0

%\cite{Ermushev:1986cu}
\bibitem{Ermushev:1986cu}
A.~V.~Ermushev, D.~I.~Kazakov and O.~V.~Tarasov,
%``FINITE N=1 SUPERSYMMETRIC GRAND UNIFIED THEORIES,''
Nucl. Phys. B \textbf{281} (1987), 72.%-84
%doi:10.1016/0550-3213(87)90247-1

%\cite{Lucchesi:1987he}
\bibitem{Lucchesi:1987he}
C.~Lucchesi, O.~Piguet and K.~Sibold,
%``Vanishing Beta Functions in $N=1$ Supersymmetric Gauge Theories,''
Helv. Phys. Acta \textbf{61} (1988), 321.
%CERN-TH-4860/87.

%\cite{Lucchesi:1987ef}
\bibitem{Lucchesi:1987ef}
C.~Lucchesi, O.~Piguet and K.~Sibold,
%``Necessary and Sufficient Conditions for All Order Vanishing Beta Functions in Supersymmetric {Yang-Mills} Theories,''
Phys. Lett. B \textbf{201} (1988), 241.%-244
%doi:10.1016/0370-2693(88)90221-3

%\cite{Parkes:1985hj}
\bibitem{Parkes:1985hj}
A.~J.~Parkes and P.~C.~West,
%``Three Loop Results in Two Loop Finite Supersymmetric Gauge Theories,''
Nucl. Phys. B \textbf{256} (1985), 340.%-352
%doi:10.1016/0550-3213(85)90397-9

%\cite{Grisaru:1985tc}
\bibitem{Grisaru:1985tc}
M.~T.~Grisaru, B.~Milewski and D.~Zanon,
%``The Structure of {UV} Divergences in Ssym Theories,''
Phys. Lett. B \textbf{155} (1985), 357.%-363
%doi:10.1016/0370-2693(85)91587-4

%\cite{Stepanyantz:2021dus}
\bibitem{Stepanyantz:2021dus}
K.~Stepanyantz,
%``Exact $\beta$-functions for ${\cal N}=1$ supersymmetric theories finite in the lowest loops,''
Eur. Phys. J. C \textbf{81} (2021), 571.
%doi:10.1140/epjc/s10052-021-09363-7
%[arXiv:2105.00900 [hep-th]].

%\cite{Jack:1995gm}
\bibitem{Jack:1995gm}
I.~Jack and D.~R.~T.~Jones,
%``Renormalization group invariance and universal soft supersymmetry breaking,''
Phys. Lett. B \textbf{349} (1995), 294.%-299
%doi:10.1016/0370-2693(95)00271-L
%[arXiv:hep-ph/9501395 [hep-ph]].

%\cite{Jack:1996qq}
\bibitem{Jack:1996qq}
I.~Jack, D.~R.~T.~Jones and C.~G.~North,
%``N=1 supersymmetry and the three loop anomalous dimension for the chiral superfield,''
Nucl. Phys. B \textbf{473} (1996), 308.%-322
%doi:10.1016/0550-3213(96)00269-6
%[arXiv:hep-ph/9603386 [hep-ph]].

%\cite{Mondragon:2013aea}
\bibitem{Mondragon:2013aea}
M.~Mondrag\'on, N.~D.~Tracas and G.~Zoupanos,
%``Reduction of Couplings in the MSSM,''
Phys. Lett. B \textbf{728} (2014), 51.
%doi:10.1016/j.physletb.2013.11.043
%[arXiv:1309.0996 [hep-ph]].

%\cite{Heinemeyer:2014vxa}
\bibitem{Heinemeyer:2014vxa}
S.~Heinemeyer, J.~Kubo, M.~Mondragon, O.~Piguet, K.~Sibold, W.~Zimmermann and G.~Zoupanos,
%``Reduction of couplings and its application in particle physics, Finite theories, Higgs and top mass predictions,''
arXiv:1411.7155 [hep-ph].

%\cite{Heinemeyer:2019vbc}
\bibitem{Heinemeyer:2019vbc}
S.~Heinemeyer, M.~Mondrag{\'o}n, N.~Tracas and G.~Zoupanos,
%``Reduction of Couplings and its application in Particle Physics,''
Phys. Rept. \textbf{814} (2019), 1.%-43
%doi:10.1016/j.physrep.2019.04.002
%[arXiv:1904.00410 [hep-ph]].

%\cite{Jack:2014pua}
\bibitem{Jack:2014pua}
I.~Jack and C.~Poole,
%``The a-function for gauge theories,''
JHEP \textbf{01} (2015), 138.
%doi:10.1007/JHEP01(2015)138
%[arXiv:1411.1301 [hep-th]].

%\cite{Kuzmichev:2023zxy}
\bibitem{Kuzmichev:2023zxy}
M.~Kuzmichev and K.~Stepanyantz,
%``A condition for the reduction of couplings in the P=13Q supersymmetric theories,''
Phys. Lett. B \textbf{844} (2023), 138094.
%doi:10.1016/j.physletb.2023.138094
%[arXiv:2306.15413 [hep-th]].

%\cite{Kutasov:2004xu}
\bibitem{Kutasov:2004xu}
D.~Kutasov and A.~Schwimmer,
%``Lagrange multipliers and couplings in supersymmetric field theory,''
Nucl. Phys. B \textbf{702} (2004), 369.%-379
%doi:10.1016/j.nuclphysb.2004.10.030
%[arXiv:hep-th/0409029 [hep-th]].

%\cite{Kataev:2013csa}
\bibitem{Kataev:2013csa}
A.~L.~Kataev and K.~V.~Stepanyantz,
%``Scheme independent consequence of the NSVZ relation for $\mathcal{N}$ = 1 SQED with $N_f$ flavors,''
Phys. Lett. B \textbf{730} (2014), 184.%-189
%doi:10.1016/j.physletb.2014.01.053
%[arXiv:1311.0589 [hep-th]].

%\cite{Kataev:2014gxa}
\bibitem{Kataev:2014gxa}
A.~L.~Kataev and K.~V.~Stepanyantz,
%``The NSVZ beta-function in supersymmetric theories with different regularizations and renormalization prescriptions,''
Theor. Math. Phys. \textbf{181} (2014), 1531.%-1540
%doi:10.1007/s11232-014-0233-3
%[arXiv:1405.7598 [hep-th]].

%\cite{Jack:1996vg}
\bibitem{Jack:1996vg}
I.~Jack, D.~R.~T.~Jones and C.~G.~North,
%``N=1 supersymmetry and the three loop gauge Beta function,''
Phys. Lett. B \textbf{386} (1996), 138.
%doi:10.1016/0370-2693(96)00918-5
%[arXiv:hep-ph/9606323 [hep-ph]].

%\cite{Jack:1996cn}
\bibitem{Jack:1996cn}
I.~Jack, D.~R.~T.~Jones and C.~G.~North,
%``Scheme dependence and the NSVZ Beta function,''
Nucl. Phys. B \textbf{486} (1997), 479.
%doi:10.1016/S0550-3213(96)00637-2
%[arXiv:hep-ph/9609325 [hep-ph]].

%\cite{Jack:1998uj}
\bibitem{Jack:1998uj}
I.~Jack, D.~R.~T.~Jones and A.~Pickering,
%``The Connection between DRED and NSVZ,''
Phys. Lett. B \textbf{435} (1998), 61.
%doi:10.1016/S0370-2693(98)00769-2
%[arXiv:hep-ph/9805482 [hep-ph]].

%\cite{Siegel:1979wq}
\bibitem{Siegel:1979wq}
W.~Siegel,
%``Supersymmetric Dimensional Regularization via Dimensional Reduction,''
Phys. Lett. B \textbf{84} (1979), 193.
%doi:10.1016/0370-2693(79)90282-X

%\cite{Bardeen:1978yd}
\bibitem{Bardeen:1978yd}
W.~A.~Bardeen, A.~J.~Buras, D.~W.~Duke and T.~Muta,
%``Deep Inelastic Scattering Beyond the Leading Order in Asymptotically Free Gauge Theories,''
Phys. Rev. D \textbf{18} (1978), 3998.
%doi:10.1103/PhysRevD.18.3998

%\cite{Siegel:1980qs}
\bibitem{Siegel:1980qs}
W.~Siegel,
%``Inconsistency of Supersymmetric Dimensional Regularization,''
Phys. Lett. B \textbf{94} (1980), 37-40
doi:10.1016/0370-2693(80)90819-9

%\cite{Avdeev:1982np}
\bibitem{Avdeev:1982np}
L.~V.~Avdeev,
%``Noninvariance of Regularization by Dimensional Reduction: An Explicit Example of Supersymmetry Breaking,''
Phys. Lett. B \textbf{117} (1982), 317.%-320
%doi:10.1016/0370-2693(82)90726-2

%\cite{Avdeev:1982xy}
\bibitem{Avdeev:1982xy}
L.~V.~Avdeev and A.~A.~Vladimirov,
%``Dimensional Regularization and Supersymmetry,''
Nucl. Phys. B \textbf{219} (1983), 262.%-276
%doi:10.1016/0550-3213(83)90437-6

%\cite{Velizhanin:2008rw}
\bibitem{Velizhanin:2008rw}
V.~N.~Velizhanin,
%``Three-loop renormalization of the N=1, N=2, N=4 supersymmetric Yang-Mills theories,''
Nucl. Phys. B \textbf{818} (2009), 95.%-100
%doi:10.1016/j.nuclphysb.2009.03.017
%[arXiv:0809.2509 [hep-th]].

%\cite{Chakraborty:2026hdv}
\bibitem{Chakraborty:2026hdv}
M.~Chakraborty and S.~O.~Moch,
%``Dimensional Reduction is Supersymmetric at Three Loops,''
arXiv:2603.02892 [hep-th].

%\cite{Kazantsev:2020kfl}
\bibitem{Kazantsev:2020kfl}
A.~Kazantsev and K.~Stepanyantz,
%``Two-loop renormalization of the matter superfields and finiteness of ${\cal N}=1$ supersymmetric gauge theories regularized by higher derivatives,''
JHEP \textbf{06} (2020), 108.
%doi:10.1007/JHEP06(2020)108
%[arXiv:2004.00330 [hep-th]].

%\cite{Slavnov:1971aw}
\bibitem{Slavnov:1971aw}
A.~A.~Slavnov,
%``Invariant regularization of nonlinear chiral theories,''
Nucl. Phys. B \textbf{31} (1971), 301.
%doi:10.1016/0550-3213(71)90234-3
%%CITATION = doi:10.1016/0550-3213(71)90234-3;%%

%\cite{Slavnov:1972sq}
\bibitem{Slavnov:1972sq}
A.~A.~Slavnov,
%``Invariant regularization of gauge theories,''
Theor. Math. Phys. \textbf{13} (1972), 1064
[Teor. Mat. Fiz. \textbf{13} (1972), 174].
%%CITATION = TMFZA,13,174;%%

%\cite{Slavnov:1977zf}
\bibitem{Slavnov:1977zf}
A.~A.~Slavnov,
%``The Pauli-Villars Regularization for Nonabelian Gauge Theories,''
Theor. Math. Phys. \textbf{33} (1977), 977
[Teor. Mat. Fiz. \textbf{33} (1977), 210].
%doi:10.1007/BF01036595
%%CITATION = doi:10.1007/BF01036595;%%

%\cite{Krivoshchekov:1978xg}
\bibitem{Krivoshchekov:1978xg}
V.~K.~Krivoshchekov,
%``Invariant Regularizations for Supersymmetric Gauge Theories,''
Theor. Math. Phys. \textbf{36} (1978), 745
[Teor. Mat. Fiz.  \textbf{36} (1978), 291].

%\cite{West:1985jx}
\bibitem{West:1985jx}
P.~C.~West,
%``Higher Derivative Regulation of Supersymmetric Theories,''
Nucl. Phys. B \textbf{268} (1986), 113.
%doi:10.1016/0550-3213(86)90203-8

%\cite{Stepanyantz:2016gtk}
\bibitem{Stepanyantz:2016gtk}
K.~V.~Stepanyantz,
%``Non-renormalization of the $V\bar cc$-vertices in ${\cal N}=1$ supersymmetric theories,''
Nucl. Phys. B \textbf{909} (2016), 316.%-335
%doi:10.1016/j.nuclphysb.2016.05.011
%[arXiv:1603.04801 [hep-th]].

%\cite{Stepanyantz:2019ihw}
\bibitem{Stepanyantz:2019ihw}
K.~V.~Stepanyantz,
%``The $\beta$-function of ${\cal N}=1$ supersymmetric gauge theories regularized by higher covariant derivatives as an integral of double total derivatives,''
JHEP \textbf{10} (2019), 011.
%doi:10.1007/JHEP10(2019)011
%[arXiv:1908.04108 [hep-th]].

%\cite{Stepanyantz:2020uke}
\bibitem{Stepanyantz:2020uke}
K.~Stepanyantz,
%``The all-loop perturbative derivation of the NSVZ $\beta$-function and the NSVZ scheme in the non-Abelian case by summing singular contributions,''
Eur. Phys. J. C \textbf{80} (2020) no.10, 911.
%doi:10.1140/epjc/s10052-020-8416-6
%[arXiv:2007.11935 [hep-th]].

%\cite{Stepanyantz:2019lyo}
\bibitem{Stepanyantz:2019lyo}
K.~Stepanyantz,
%``The Higher Covariant Derivative Regularization as a Tool for Revealing the Structure of Quantum Corrections in Supersymmetric Gauge Theories,''
Proc. Steklov Inst. Math. \textbf{309} (2020) no.1, 284.%-298
%doi:10.1134/S0081543820030219
%[arXiv:1910.03242 [hep-th]].

%\cite{Stepanyantz:2023jot}
\bibitem{Stepanyantz:2023jot}
K.~V.~Stepanyantz,
%``The structure of quantum corrections and exact results in supersymmetric theories from the higher covariant derivative regularization,''
Theor. Math. Phys. \textbf{217} (2023) no.3, 1954.%-1968
%doi:10.1134/S0040577923120127

%\cite{Kataev:2013eta}
\bibitem{Kataev:2013eta}
A.~L.~Kataev and K.~V.~Stepanyantz,
%``NSVZ scheme with the higher derivative regularization for $\mathcal{N} =$ 1 SQED,''
Nucl. Phys. B \textbf{875} (2013), 459.%-482
%doi:10.1016/j.nuclphysb.2013.07.010
%[arXiv:1305.7094 [hep-th]].

%\cite{Kataev:2019olb}
\bibitem{Kataev:2019olb}
A.~L.~Kataev, A.~E.~Kazantsev and K.~V.~Stepanyantz,
%``On-shell renormalization scheme for ${{\mathcal {N}}}=1$ SQED and the NSVZ relation,''
Eur. Phys. J. C \textbf{79} (2019) no.6, 477.
%doi:10.1140/epjc/s10052-019-6993-z
%[arXiv:1905.02222 [hep-th]].

%\cite{Stepanyantz:2017sqg}
\bibitem{Stepanyantz:2017sqg}
K.~V.~Stepanyantz,
%``Structure of Quantum Corrections in ${\cal N}=1$ Supersymmetric Gauge Theories,''
Bled Workshops Phys. \textbf{18} (2017) no.2, 197.%-213
%[arXiv:1711.09194 [hep-th]].

%\cite{Shakhmanov:2017wji}
\bibitem{Shakhmanov:2017wji}
V.~Y.~Shakhmanov and K.~V.~Stepanyantz,
%``New form of the NSVZ relation at the two-loop level,''
Phys. Lett. B \textbf{776} (2018), 417.
%doi:10.1016/j.physletb.2017.12.005
%[arXiv:1711.03899 [hep-th]].

%\cite{Stepanyantz:2011jy}
\bibitem{Stepanyantz:2011jy}
K.~V.~Stepanyantz,
%``Derivation of the exact NSVZ $\beta$-function in N=1 SQED, regularized by higher derivatives, by direct summation of Feynman diagrams,''
Nucl. Phys. B \textbf{852} (2011), 71.%-107
%doi:10.1016/j.nuclphysb.2011.06.018
%[arXiv:1102.3772 [hep-th]].

%\cite{Kataev:2017qvk}
\bibitem{Kataev:2017qvk}
A.~L.~Kataev, A.~E.~Kazantsev and K.~V.~Stepanyantz,
%``The Adler $D$-function for ${\cal N}=1$ SQCD regularized by higher covariant derivatives in the three-loop approximation,''
Nucl. Phys. B \textbf{926} (2018), 295.%-320
%doi:10.1016/j.nuclphysb.2017.11.009
%[arXiv:1710.03941 [hep-th]].

%\cite{Shakhmanov:2017soc}
\bibitem{Shakhmanov:2017soc}
V.~Y.~Shakhmanov and K.~V.~Stepanyantz,
%``Three-loop NSVZ relation for terms quartic in the Yukawa couplings with the higher covariant derivative regularization,''
Nucl. Phys. B \textbf{920} (2017), 345.%-367
%doi:10.1016/j.nuclphysb.2017.04.017
%[arXiv:1703.10569 [hep-th]].

%\cite{Kazantsev:2018nbl}
\bibitem{Kazantsev:2018nbl}
A.~E.~Kazantsev, V.~Y.~Shakhmanov and K.~V.~Stepanyantz,
%``New form of the exact NSVZ $\beta$-function: the three-loop verification for terms containing Yukawa couplings,''
JHEP \textbf{04} (2018), 130.
%doi:10.1007/JHEP04(2018)130
%[arXiv:1803.06612 [hep-th]].

%\cite{Kuzmichev:2019ywn}
\bibitem{Kuzmichev:2019ywn}
M.~D.~Kuzmichev, N.~P.~Meshcheriakov, S.~V.~Novgorodtsev, I.~E.~Shirokov and K.~V.~Stepanyantz,
%``Three-loop contribution of the Faddeev\textendash{}Popov ghosts to the $\beta $ -function of $\mathcal{N}=1$ supersymmetric gauge theories and the NSVZ relation,''
Eur. Phys. J. C \textbf{79} (2019) no.9, 809.
%doi:10.1140/epjc/s10052-019-7323-1
%[arXiv:1908.10586 [hep-th]].

%\cite{Aleshin:2020gec}
\bibitem{Aleshin:2020gec}
S.~S.~Aleshin, \textit{et al.}
%``Three-loop verification of a new algorithm for the calculation of a $\beta$-function in supersymmetric theories regularized by higher derivatives for the case of ${\cal N}=1$ SQED,''
Nucl. Phys. B \textbf{956} (2020), 115020.
%doi:10.1016/j.nuclphysb.2020.115020
%[arXiv:2003.06851 [hep-th]].

%\cite{Shirokov:2023jya}
\bibitem{Shirokov:2023jya}
I.~Shirokov and V.~Shirokova,
%``The four-loop $\beta $-function from vacuum supergraphs and the NSVZ relation for ${{{\mathcal {N}}}}=1$ SQED regularized by higher derivatives,''
Eur. Phys. J. C \textbf{84} (2024) no.3, 249.
%doi:10.1140/epjc/s10052-024-12587-y
%[arXiv:2310.13109 [hep-th]].

%\cite{Shirokov:2022jyd}
\bibitem{Shirokov:2022jyd}
I.~Shirokov and K.~Stepanyantz,
%``The three-loop anomalous dimension and the four-loop {\ensuremath{\beta}}-function for $ \mathcal{N} $ = 1 SQED regularized by higher derivatives,''
JHEP \textbf{04} (2022), 108.
%doi:10.1007/JHEP04(2022)108
%[arXiv:2203.01113 [hep-th]].

%\cite{Haneychuk:2022qvu}
\bibitem{Haneychuk:2022qvu}
O.~Haneychuk, V.~Shirokova and K.~Stepanyantz,
%``Three-loop \ensuremath{\beta}-functions and two-loop anomalous dimensions for MSSM regularized by higher covariant derivatives in an arbitrary supersymmetric subtraction scheme,''
JHEP \textbf{09} (2022), 189.
%doi:10.1007/JHEP09(2022)189
%[arXiv:2207.11944 [hep-ph]].

%\cite{Haneychuk:2025ejd}
\bibitem{Haneychuk:2025ejd}
O.~Haneychuk,
%``General Expression for Three-Loop {\ensuremath{\beta}}-Functions of $\mathcal{N}{\mathbf{ = 1}}$ Supersymmetric Theories with Multiple Gauge Couplings Regularized by Higher Covariant Derivatives,''
JETP Lett. \textbf{121} (2025) no.5, 320.%-323
%doi:10.1134/S0021364025600119
%[arXiv:2501.05174 [hep-th]].

%\cite{Jack:1997sr}
\bibitem{Jack:1997sr}
I.~Jack and D.~R.~T.~Jones,
%``Regularization of supersymmetric theories,''
Adv. Ser. Direct. High Energy Phys. \textbf{21} (2010), 494.%-513
%doi:10.1142/9789814307505{\_}0013
%[arXiv:hep-ph/9707278 [hep-ph]].

%\cite{Gnendiger:2017pys}
\bibitem{Gnendiger:2017pys}
C.~Gnendiger, A.~Signer, D.~St\"ockinger, A.~Broggio, A.~L.~Cherchiglia, F.~Driencourt-Mangin, A.~R.~Fazio, B.~Hiller, P.~Mastrolia and T.~Peraro, \textit{et al.}
%``To ${d}$, or not to ${d}$: recent developments and comparisons of regularization schemes,''
Eur. Phys. J. C \textbf{77} (2017) no.7, 471.
%doi:10.1140/epjc/s10052-017-5023-2
%[arXiv:1705.01827 [hep-ph]].

%\cite{Arkani-Hamed:1997qui}
\bibitem{Arkani-Hamed:1997qui}
N.~Arkani-Hamed and H.~Murayama,
%``Holomorphy, rescaling anomalies and exact beta functions in supersymmetric gauge theories,''
JHEP \textbf{06} (2000), 030.
%doi:10.1088/1126-6708/2000/06/030
%[arXiv:hep-th/9707133 [hep-th]].

%\cite{Stepanyantz:2004sg}
\bibitem{Stepanyantz:2004sg}
K.~V.~Stepanyantz,
%``Investigation of the anomaly puzzle in N = 1 supersymmetric electrodynamics,''
Theor. Math. Phys. \textbf{142} (2005), 29.%-47
%doi:10.1007/s11232-005-0053-6
%[arXiv:hep-th/0407201 [hep-th]].

%\cite{Novikov:1985mf}
\bibitem{Novikov:1985mf}
V.~A.~Novikov, M.~A.~Shifman, A.~I.~Vainshtein and V.~I.~Zakharov,
%``Supersymmetric Extension of the Adler-bardeen Theorem,''
Phys. Lett. B \textbf{157} (1985), 169.%-173
%doi:10.1016/0370-2693(85)91539-4

%\cite{Goriachuk:2018cac}
\bibitem{Goriachuk:2018cac}
I.~O.~Goriachuk, A.~L.~Kataev and K.~V.~Stepanyantz,
%``A class of the NSVZ renormalization schemes for ${\cal N}=1$ SQED,''
Phys. Lett. B \textbf{785} (2018), 561.%-566
%doi:10.1016/j.physletb.2018.09.014
%[arXiv:1808.02050 [hep-th]].

\bibitem{Goriachuk_Conference}
I.~O.~Goriachuk,
%``A class of the NSVZ schemes in supersymmetric gauge theories'',
Proceedings of XXVI International conference of students, graduate students, and young scientists on fundamental sciences ``Lomonosov--2019'', section ``Physics'' (2019),
https://istina.msu.ru/download/382190943/1m2xDG:K-PTE0Np2rtbCDO7R7N6lK0BOc4/.

%\cite{Haneychuk:2025ehb}
\bibitem{Haneychuk:2025ehb}
O.~Haneychuk and K.~Stepanyantz,
%``Three-loop verification of the equations relating running of the gauge couplings in $\mathcal{N}=1$ SQCD + SQED,''
Eur. Phys. J. C \textbf{85} (2025) no.5, 540.
%doi:10.1140/epjc/s10052-025-14250-6
%[arXiv:2501.06500 [hep-th]].

%\cite{Gates:1983nr}
\bibitem{Gates:1983nr}
S.~J.~Gates, M.~T.~Grisaru, M.~Rocek and W.~Siegel,
%``Superspace Or One Thousand and One Lessons in Supersymmetry,''
Front. Phys. \textbf{58} (1983), 1.
%[arXiv:hep-th/0108200 [hep-th]].

%\cite{West:1990tg}
\bibitem{West:1990tg}
P.~C.~West,
``Introduction to supersymmetry and supergravity,''
Singapore, Singapore: World Scientific (1990) 425 p.

%\cite{Buchbinder:1998qv}
\bibitem{Buchbinder:1998qv}
I.~L.~Buchbinder and S.~M.~Kuzenko,
``Ideas and methods of supersymmetry and supergravity: Or a walk through superspace,''
Bristol, UK: IOP (1998), 656 p.

%\cite{Piguet:1981fb}
\bibitem{Piguet:1981fb}
O.~Piguet and K.~Sibold,
%``Renormalization of $N=1$ Supersymmetrical {Yang-Mills} Theories. 1. The Classical Theory,''
Nucl. Phys. B \textbf{197} (1982), 257.%-271
%doi:10.1016/0550-3213(82)90291-7

%\cite{Piguet:1981hh}
\bibitem{Piguet:1981hh}
O.~Piguet and K.~Sibold,
%``Renormalization of $N=1$ Supersymmetrical {Yang-Mills} Theories. 2. The Radiative Corrections,''
Nucl. Phys. B \textbf{197} (1982), 272.%-289
%doi:10.1016/0550-3213(82)90292-9

%\cite{Tyutin:1983rg}
\bibitem{Tyutin:1983rg}
I.~V.~Tyutin,
%``RENORMALIZATION OF SUPERGAUGE THEORIES WITH NONEXTENDED SUPERSYMMETRY. (IN RUSSIAN),''
Yad. Fiz. \textbf{37} (1983), 761.%-771

%\cite{Juer:1982fb}
\bibitem{Juer:1982fb}
J.~W.~Juer and D.~Storey,
%``Nonlinear Renormalization in Superfield Gauge Theories,''
Phys. Lett. B \textbf{119} (1982), 125.%-127
%doi:10.1016/0370-2693(82)90259-3

%\cite{Juer:1982mp}
\bibitem{Juer:1982mp}
J.~W.~Juer and D.~Storey,
%``One Loop Renormalization of Superfield {Yang-Mills} Theories,''
Nucl. Phys. B \textbf{216} (1983), 185.%-208
%doi:10.1016/0550-3213(83)90491-1

%\cite{Kazantsev:2018kjx}
\bibitem{Kazantsev:2018kjx}
A.~E.~Kazantsev, M.~D.~Kuzmichev, N.~P.~Meshcheriakov, S.~V.~Novgorodtsev, I.~E.~Shirokov, M.~B.~Skoptsov and K.~V.~Stepanyantz,
%``Two-loop renormalization of the Faddeev-Popov ghosts in $ \mathcal{N}=1 $ supersymmetric gauge theories regularized by higher derivatives,''
JHEP \textbf{06} (2018), 020.
%doi:10.1007/JHEP06(2018)020
%[arXiv:1805.03686 [hep-th]].

%\cite{Aleshin:2016yvj}
\bibitem{Aleshin:2016yvj}
S.~S.~Aleshin, A.~E.~Kazantsev, M.~B.~Skoptsov and K.~V.~Stepanyantz,
%``One-loop divergences in non-Abelian supersymmetric theories regularized by BRST-invariant version of the higher derivative regularization,''
JHEP \textbf{05} (2016), 014.
%doi:10.1007/JHEP05(2016)014
%[arXiv:1603.04347 [hep-th]].

%\cite{Kazantsev:2017fdc}
\bibitem{Kazantsev:2017fdc}
A.~E.~Kazantsev, M.~B.~Skoptsov and K.~V.~Stepanyantz,
%``One-loop polarization operator of the quantum gauge superfield for ${\cal N}=1$ SYM regularized by higher derivatives,''
Mod. Phys. Lett. A \textbf{32} (2017) no.36, 1750194.
%doi:10.1142/S0217732317501942
%[arXiv:1709.08575 [hep-th]].

%\cite{Singh:2025fgo}
\bibitem{Singh:2025fgo}
S.~K.~Singh,
%``Three-Loop Gauge Beta Functions in Supersymmetric Theories with Exponential Higher Covariant Derivative Regularization,''
arXiv:2509.06799 [hep-th].

%\cite{Shifman:1996iy}
\bibitem{Shifman:1996iy}
M.~A.~Shifman,
%``Little miracles of supersymmetric evolution of gauge couplings,''
Int. J. Mod. Phys. A \textbf{11} (1996), 5761.
%doi:10.1142/S0217751X96002650
%[arXiv:hep-ph/9606281 [hep-ph]].

%\cite{Korneev:2021zdz}
\bibitem{Korneev:2021zdz}
D.~Korneev, D.~Plotnikov, K.~Stepanyantz and N.~Tereshina,
%``The NSVZ relations for $ \mathcal{N} $ = 1 supersymmetric theories with multiple gauge couplings,''
JHEP \textbf{10} (2021), 046.
%doi:10.1007/JHEP10(2021)046
%[arXiv:2108.05026 [hep-th]].

%\cite{Vainshtein:1985ynw}
\bibitem{Vainshtein:1985ynw}
A.~I.~Vainshtein, V.~I.~Zakharov and M.~A.~Shifman,
%``GELL-MANN-LOW FUNCTION IN SUPERSYMMETRIC ELECTRODYNAMICS,''
JETP Lett. \textbf{42} (1985), 224.%-227

%\cite{Shifman:1985fi}
\bibitem{Shifman:1985fi}
M.~A.~Shifman, A.~I.~Vainshtein and V.~I.~Zakharov,
%``EXACT GELL-MANN-LOW FUNCTION IN SUPERSYMMETRIC ELECTRODYNAMICS,''
Phys. Lett. B \textbf{166} (1986), 334.
%doi:10.1016/0370-2693(86)90811-7

%\cite{Smilga:2004zr}
\bibitem{Smilga:2004zr}
A.~V.~Smilga and A.~Vainshtein,
%``Background field calculations and nonrenormalization theorems in 4-D supersymmetric gauge theories and their low-dimensional descendants,''
Nucl. Phys. B \textbf{704} (2005), 445.%-474
%doi:10.1016/j.nuclphysb.2004.10.010
%[arXiv:hep-th/0405142 [hep-th]].

%\cite{Kazantsev:2014yna}
\bibitem{Kazantsev:2014yna}
A.~E.~Kazantsev and K.~V.~Stepanyantz,
%``Relation between two-point Green{\textquoteright}s functions of $\mathcal{N} = 1$ SQED with N$_{f}$ flavors, regularized by higher derivatives, in the three-loop approximation,''
J. Exp. Theor. Phys. \textbf{120} (2015) no.4, 618.%-631
%doi:10.1134/S1063776115040068
%[arXiv:1410.1133 [hep-th]].

%\cite{Lakhal:2025nbh}
\bibitem{Lakhal:2025nbh}
A.~Lakhal and K.~Stepanyantz,
%``Relation between leading divergences in nonrenormalizable 4D supersymmetric theories,''
Eur. Phys. J. C \textbf{86} (2026) no.3, 313.
%doi:10.1140/epjc/s10052-026-15530-5
%[arXiv:2512.12780 [hep-th]].

%\cite{Rosner:1967zz}
\bibitem{Rosner:1967zz}
J.~L.~Rosner,
%``Higher-order contributions to the divergent part of Z(3) in a model quan tum electrodynamics,''
Annals Phys. \textbf{44} (1967), 11.
%doi:10.1016/0003-4916(67)90262-X

%\cite{Derkachev:2017nhd}
\bibitem{Derkachev:2017nhd}
C.~E.~Derkachev, A.~V.~Ivanov and L.~D.~Faddeev,
%``Renormalization scenario for the quantum Yang\textendash{}Mills theory in four-dimensional space\textendash{}time,''
Theor. Math. Phys. \textbf{192} (2017) no.2, 1134 [Teor. Mat. Fiz. \textbf{192} (2017) no.2, 227].
%doi:10.1134/S0040577917080049

%\cite{Meshcheriakov:2022tyi}
\bibitem{Meshcheriakov:2022tyi}
N.~Meshcheriakov, V.~Shatalova and K.~Stepanyantz,
%``Coefficients at powers of logarithms in the higher-derivatives and minimal-subtractions-of-logarithms renormalization scheme,''
Phys. Rev. D \textbf{106} (2022) no.10, 105011.
%doi:10.1103/PhysRevD.106.105011
%[arXiv:2208.13443 [hep-th]].

%\cite{Meshcheriakov:2023fmk}
\bibitem{Meshcheriakov:2023fmk}
N.~Meshcheriakov, V.~Shatalova and K.~Stepanyantz,
%``Higher logarithms and {\ensuremath{\varepsilon}}-poles for the MS-like renormalization prescriptions,''
JHEP \textbf{12} (2023), 097.
%doi:10.1007/JHEP12(2023)097
%[arXiv:2310.05610 [hep-th]].

%\cite{Meshcheriakov:2024qwj}
\bibitem{Meshcheriakov:2024qwj}
N.~Meshcheriakov, V.~Shatalova and K.~Stepanyantz,
%``Higher ${\varepsilon}$-Poles and Logarithms in the MS-Like Schemes from the Algebraic Structure of the Renormalization Group,''
Moscow Univ. Phys. Bull. \textbf{80} (2025) no.4, 664.%-675
%doi:10.3103/S0027134925700833
%[arXiv:2405.11557 [hep-th]].

%\cite{Kovyrshin:2025ufp}
\bibitem{Kovyrshin:2025ufp}
G.~Kovyrshin, N.~Meshcheriakov, V.~Shatalova and K.~Stepanyantz,
%``Structure of renormalization constants for theories with multiple couplings in the MS-like subtraction schemes,''
Nucl. Phys. B \textbf{1022} (2026), 117272.
%doi:10.1016/j.nuclphysb.2025.117272
%[arXiv:2509.03437 [hep-th]].

%\cite{Harlander:2006xq}
\bibitem{Harlander:2006xq}
R.~V.~Harlander, D.~R.~T.~Jones, P.~Kant, L.~Mihaila and M.~Steinhauser,
%``Four-loop beta function and mass anomalous dimension in dimensional reduction,''
JHEP \textbf{12} (2006), 024.
%doi:10.1088/1126-6708/2006/12/024
%[arXiv:hep-ph/0610206 [hep-ph]].

\end{thebibliography}
\end{document}